\documentclass[12pt,twocolumn,preprint]{iopart}
\usepackage{iopams}  
\usepackage{graphicx}
\usepackage{subfig}
\begin{document}
 
\title{Evaluation of ideal MHD mode stability of CFETR baseline scenario}

\author{Debabrata Banerjee}

\address{CAS Key Laboratory of Geospace Environment and Department of Modern 
Physics, University of Science and Technology of China, Hefei, Anhui 230026, 
China}

\author{Ping Zhu}

\address{CAS Key Laboratory of Geospace Environment and Department of Modern 
Physics, University of Science and Technology of China, Hefei, Anhui 230026, China}
\address{KTX Laboratory and Department of Modern Physics, University of Science and Technology of China, Hefei, Anhui 230026, China}
\address{Department of Engineering Physics, University of Wisconsin-Madison, Madison, Wisconsin 53706, USA}
\ead{pzhu@ustc.edu.cn}

\author{Shikui Cheng}

\address{CAS Key Laboratory of Geospace Environment and Department of Modern 
Physics, University of Science and Technology of China, Hefei, Anhui 230026, 
China}

\author{Xingting Yan}

\address{CAS Key Laboratory of Geospace Environment and Department of Modern 
Physics, University of Science and Technology of China, Hefei, Anhui 230026, 
China}

\author{Rui Han}

\address{CAS Key Laboratory of Geospace Environment and Department of Modern 
Physics, University of Science and Technology of China, Hefei, Anhui 230026, 
China}

\author{Linjin Zheng}%

\address{Institute of Fusion Studies, University of Texas at Austin, Texas, 78712, USA}

\author{The CFETR Physics Team}

\newpage

\begin{abstract}
The CFETR baseline scenario is based on a H-mode equilibrium with high pedestal and highly peaked edge bootstrap current, along with strong reverse shear in safety factor profile.
The stability of ideal MHD modes for the CFETR baseline scenario has been evaluated using NIMROD and AEGIS codes. The toroidal mode numbers (n=1-10) are considered in this analysis 
for different positions of perfectly conducting wall in order to estimate the ideal wall effect on the stability of ideal MHD modes for physics and engineering designs of CFETR. 
Although, the modes (n=1-10) are found to be unstable in ideal MHD, the structure of all modes is edge localized. Growth rates of all modes are found to be increasing initially with 
wall position before they reach ideal wall saturation limit (no wall limit). No global core modes are found to be dominantly unstable in our analysis. The design of $q_{min}>2$ and strong reverse 
shear in $q$ profile is expected to prevent the excitation of global modes. Therefore, this baseline scenario is considered to be suitable for supporting long time steady state discharge 
in context of ideal MHD physics, if ELMs could be controlled. 
\end{abstract}

%
%
%
\maketitle
%
%
\section{Introduction}
Besides being a partner in ITER~\cite{aymer}, China has recently proposed to design and potentially build China Fusion Engineering 
Test Reactor~(CFETR)~\cite{wan2014}. The goal is to address the physics and engineering issues essential 
for bridging the gap between ITER and DEMO, and to promote the advancement towards fusion reactor. These issues include efficient breeding of tritium after capturing 
high energy neutrons into lithium blanket and exploring option for DEMO blanket and divertor solutions. 
CFETR is expected to achieve high annual duty factor of $0.3-0.5$ and demonstrate tritium self-sufficiency with target tritium breeding rate greater than $1$~\cite{chan2015,shi2016}.

A conceptual engineering design of CFETR including different coils and remote maintenance systems
was prepared in the beginning~\cite{song2014}. The initial parameters of CFETR was set up through running a 0-D system code~\cite{wan2014}, and later these are optimized involving
different other system codes (GASC and TESC)~\cite{chan2015}. The preliminary design of snowflake divertor for CFETR has been made, and simulation work is carried
out to evaluate heat flux onto the divertor~\cite{mao2015}. To fulfill different physics goals, the CFETR has been designed for two steady-state scenarios - baseline and advanced scenarios. 
Baseline CFETR scenario is designed to achieve moderate fusion power (200 MW) applying a fully non-inductive current drive, giving more importance towards challenging annual duty factor $0.3-0.5$. 
So, the idea is to achieve these targets with a conservative stable physics scenario first, before finally moving to the advanced scenario. Advanced design is aimed at
higher fusion power and gain close to fusion reactor with challenging fraction of non-inductive bootstrap current drive. A detailed comparison between these two scenarios using different system codes 
analysis has been reported in a recent article~\cite{shi2016}

Due to the goal of achieving high $\beta$ and high fraction of non-inductive bootstrap current in CFETR, both pressure and current driven instabilities are likely to threaten steady state operation. 
To confirm the viability of long duration steady state operation in CFETR scenarios, a thorough evaluation of the stability of all ideal MHD modes is essential, so that a stable parameter space could be determined. 
The strong reverse shear in safety factor profile and the optimized design of $q_{min}>2$ are expected to stabilize different devastating global core modes, such as $(1,1)$ and $(2,1)$ internal kink modes. 
The requirement of moderate to high fusion power gain in CFETR, would require higher pedestal top pressure value resulting in a steeper gradient in edge pressure profile. The aim for 
fully non-inductive operation, has proposed requirement of $36\%$ and $74\%$ of bootstrap current fraction to baseline and advanced scenarios respectively, whereas the ITER steady state is designed to be $48\%$ (see Table-1 of \cite{xian2017}).  These 
requirements lead to high pedestal $\beta$ and peaked edge current, which are expected to drive the excitation of edge localized modes (ELMs). The repetitive expulsion of stored plasma energy and particles due to ELMs, would degrade plasma confinement and damage
divertor and first wall components. For reactor scale machines, the sizes of ELMs are projected to be larger than those in current tokamaks~\cite{loarte2007,leonard}. Thus, stability analysis of ELMs is essential for further evaluating and optimizing
the design of CFETR baseline scenario. 

The present article reports the  stability analysis of the ideal MHD modes for CFETR baseline scenarios using the initial value extended-MHD code NIMROD~\cite{sovinec2004} and the eigen-value code AEGIS~\cite{zheng2006}.
In the ideal MHD model, the stability of $n=1-3$ modes are evaluated using both NIMROD and AEGIS codes, and the growth rates are compared. Also, in another calculation, we have used Spitzer resistivity to represent realistic resistive regimes of CFETR scenarios. 
The effect of conducting wall on the growth rates of $n=1-10$ modes has been studied with different positions of CFETR wall. The objective is to find the no wall limit of ideal mode growth rates and to provide physics base for the engineering design on the optimal 
choice of wall position of CFETR. 

The rest of this paper is organized as follows. In the second section, the equilibrium profiles of baseline scenario are introduced. In the third section, the resistive single-fluid MHD model in the NIMROD code is described. 
In the first subsection of the fourth section, the ideal MHD results from NIMROD is described  with the benchmarking  between NIMROD and AEGIS codes. In the second subsection of the fourth section, the influence of Spitzer profile on the stability of ideal MHD modes are shown. 
Finally, the main points are summarized and the conclusion is drawn.

\section{Equilibrium of CFETR Baseline Scenario} 
We consider the equilibrium of CFETR baseline scenario in our calculation. The necessary physics and engineering parameters of this scenario was first set up through 0-D system code analysis. 
Then, this equilibrium has self-consistently been generated through multi-dimensional integrated modeling in OMFIT framework using the auxiliary heating source in a combination of electron cyclotron wave
and neutral beam injection~\cite{xian2017}. The plasma size is slightly smaller than ITER, with a major radius of $5.7$ m and a minor radius 
of $1.6$ m. The toroidal magnetic field ($5$T) and the plasma current (10 MA) at magnetic axis are listed in Table~1 of reference~\cite{xian2017}, among others. Since the baseline case is not designed for demonstrating high fusion gain, the normalized 
$\beta_N$ is set to be $1.8$, well below the no-wall $\beta$ limit $\beta_N \sim 4 \times l_{i}$ where $l_i$ is the plasma inductance. This is expected to help this equilibrium to lie within stability limits of ideal MHD global modes. 
The plasma profiles of electron number density, ion temperature, safety factor and current density are shown as functions of square root of the normalized poloidal magnetic flux. Both density (Fig.~1a) and temperature (Fig.~1b) profiles 
show an edge pedestal region inside LCFS.  Safety factor (q) profile has strong reverse shear region (Fig.~1c) and 
 $q_{min}>2$ with low core current in order to avoid sawtooth crash. The current density profile  has highly peaked edge current due to high fraction of bootstrap current (Fig.~1d).

\section{Magneto-hydrodynamic (MHD) Model in NIMROD and AEGIS}
The MHD equations used in our NIMROD calculations are: 
\begin{eqnarray}
\frac{\partial n}{\partial t} + \nabla \cdot \left( n {\bf u}\right) = 0 \\
m n \left( \frac{\partial}{\partial t} + {\bf u} \cdot \nabla \right) {\bf u} = {\bf J} \times {\bf B} - \nabla p - \nabla \cdot \overline{\Pi} \\
\frac{3}{2} n \left( \frac{\partial}{\partial t} + {\bf u} \cdot \nabla \right) T = -n T_{\alpha} \nabla \cdot {\bf u}_{\alpha} (\alpha=i,e) \\
\frac{\partial {\bf B}}{\partial t} = - \nabla \times \left[ \eta {\bf J} - {\bf u}\times{\bf B}\right] \\ 
\mu_0 {\bf J} = \nabla \times {\bf B} ; ~~~~~~~~~~~~~~~ \nabla \cdot {\bf B} = 0
\end{eqnarray}
where {\bf u} is the center-of-mass flow velocity with particle density $n$ and ion mass $m$, $p$ is the combined pressure of electron ($p_e$) and ion ($p_i$), $\eta$ represents resistivity, 
and $\overline{\Pi}$ is the ion viscous stress tensor. The initial value NIMROD code has been consistently used in studying different macroscopic phenomena in both fusion and space plasmas~\cite{burke2010,king2016,zhu2013}.

The AEGIS code solves ideal MHD eigen-value equation employing adaptive shooting method along radial direction and Fourier decomposition in poloidal and toroidal direction.
This code has been efficiently used before in evaluating stability of low-$n$ modes in presence of both conducting and resistive walls~\cite{zheng2006,zheng2005}.
In AEGIS, ideal MHD formalism has been used to evaluate linear stability of toroidal modes $n=1-3$, where the plasma region within separatrix is modeled to have zero resistivity, and the vacuum region extended from separatrix to 
conducting wall, does not contain any plasma or current, amounting to infinite resistivity. On the contrary, NIMROD uses the resistive MHD model for both the hot core plasma within separatrix and the low density, low temperature plasma of
vacuum-like halo region between separatrix and conducting wall. So, for the purpose of comparison with the ideal MHD results, a hyperbolic tangent resistiy profile is adopted in NIMROD to represent the lowly resistive core plasma and highly
resistive vacuum region. Employing this resistivity model, a comparison in growth rates is drawn between NIMROD and AEGIS results for the $n=1-3$ modes.

%

\section{Results of Ideal MHD Stability Analysis}

\subsection{Step function profile of resistivity}

\subsubsection{Ideal MHD stability analysis in NIMROD} ~~\\

The dimensionless parameter to model the perfectly conducting ideal core plasma and infinitely resistive vacuum-like region is the Lundquist number 
defined as $S=\tau_R/\tau_A$, where resistive diffusion time $\tau_R = \mu_0 a^2/\eta$ with $\mu_0$ being the permeability of free space, $\eta$ resistivity, $a$ minor radius and Alfven time $\tau_A=R_0\sqrt{\mu_0\rho_{m0}}/B_0$ with $R_0$  being the radius of the magnetic axis, $B_0$ and $\rho_{m0}$  the values of 
magnetic field and mass density at magnetic axis respectively. The profile of Lundquist number (inverse of resistivity) is specified as a function of the normalized poloidal flux with step-like hyperbolic tangent form shown in Fig.~2, where the Lundquist numbers in plasma and vacuum regions 
are denoted as $S_{plasma}$ and $S_{vac}$ respectively. Following the same procedure described in earlier references~\cite{burke2010,debunf}, $S_{plasma}$ was scanned to find its value in the ideal MHD regime. The value of $S_{plasma}/S_{vac}$ is set to be $10^{10}/10^1$ and then growth rates of toroidal modes  $n=1-10$ are calculated with conducting wall
located at $1.2a$, where $a$ is the minor radius of plasma. As shown in Fig.~3a, the growth rate (normalized in Alfven time) of toroidal mode increases with mode number $n$, with the fastest growing one being $n=10$.  

Then growth rates of all these modes are calculated after varying the position of conducting wall in a wide range starting from close to LCFS to a wall distance from magnetic axis at $b=1.8a$. The growth rates of $n=1,3,5,8$ reach the no-wall limit at close proximity of LCFS, which indicates that the ideal MHD growth rate 
for this baseline case does not depend much on the conducting wall position (Fig. 3b). Perturbed pressure and radial magnetic field for $n=1,8$ are edge localized  at the edge pedestal near separatrix as shown in Fig.~4 (the location of
separatrix is indicated by black lines of poloidal flux contour). From Fig.~4b, the poloidal mode structure has poloidal mode number $m=4$ for $n=1$, and thus the rational surface can be identified as $q=4$ which locates at the pedestal. An apparent difference in mode structure between the $n=1$ mode and the 
$n=8$ mode is noticeable from Fig.~4, where the $n=1$ mode has broader radial structure than $n=8$.   

\subsubsection{Ideal MHD stability analysis in AEGIS and comparison with NIMROD}~~\\

A comparison between NIMROD and AEGIS results is performed for modes $n=1-3$. The ideal MHD growth rates of $n=1-3$ modes have been evaluated using AEGIS code for the same equilibrium discussed in Section~2. The comparison of normalized growth rates is shown in the Fig.~5 for two different wall locations at $b=1.35a$ and $1.5a$. 
It is clear that $n=2$ has good agreement in growth rate between NIMROD and AEGIS. Modes $n=1,3$ have slight differences in growth rates between these codes. Perturbed radial displacements of mode $n=2$ calculated in AEGIS are plotted in Fig.~6 for 
different poloidal harmonics. Both real and imaginary part of these eigenfunctions have one harmonic to be external kink mode  and others may have internal mode structures peaked around rational surfaces. 


\subsection{Spitzer model profile of resistivity}

\subsubsection{Stabilizing role of resistivity profile}~~\\

The stability of modes $n=1-10$ has been re-calculated after considering the Spitzer resisitivity profile that is  $\eta(T_e)=\eta_0 (T_{e0}/T_{e})^{3/2}$, where $T_{e0}$, $\eta_0$, $T_e$ denote the electron temperature, resistivity at magnetic axis, and the
electron temperature profile respectively. Now, our equilibrium configuration has a resisitivity profile covering whole simulation domain depending on the radial profile of electron temperature. 
The inclusion of resistivity profile is expected to make the numerical modeling more accurate for predicting the stability of CFETR baseline design. In a recent article, resistivity has been reported to have both stabilizing and destabilizing effects on ideal MHD edge localized modes~\cite{debupop}. 
The linear calculation in NIMROD has checked the stability of modes $n=1-10$ after placing conducting wall at $1.2a$ and found modes $n=2-10$ to be unstable, where the $n=1$ mode is 
always stable (blue curve in Fig.~7b). Presence of Spitzer resistivity profile leads to lower the growth rates of $n=2-10$ modes and the stabilization of the $n=1$ mode as 
compared to the results shown in Fig.~3a using the hyperbolic tangent profile. This result is consistent with earlier studies using NIMROD in the context of other tokamaks equilibria such as NSTX and JT-60U~\cite{king2016,debunf}. 

\subsubsection{Growth rate variation with wall position and shape}~~\\

The effect of conducting wall position on growth rate of all modes has been evaluated after considering self-similar wall configuration and Spitzer resistivity profile.   
The wall position has been varied in the calculation until the no wall limit of growth rate is reached. The normalized growth rates of $n=2,3,5,8,10$ modes are plotted in the Fig.~7a with  
wall position changing from close to separatrix to $b=1.8a$. The stable position of conducting wall is found to be at $1.04a$, a little away from plasma boundary. No mode is found unstable inside this position of wall.  
The growth rates of all modes vary with wall position in a similar way. Initially, it increases rapidly until the wall position $1.2a$ is reached. Afterwards, it gradually approaches the no-wall limit value. 
While the wall positions for all modes transitioning to no-wall limit are basically same, the no wall limit growth rate increases monotonically with mode numbers from $n=2$ to $n=10$.  

The results in previous paragraph are calculated for self-similar wall. The growth rate calculations of different modes have also been carried out after considering recently proposed real shaped wall configuration of CFETR. The present wall position is near to the wall location of $b=1.2a$, but shape is different from regular self-similar wall.
A clear stabilizing effect of real shape of wall is found as compared to the self-similar wall at $b=1.2a$~(Fig.~7b). The growth rates of $n=1-10$ for two different wall positions $b=1.08a,1.2a$ with self-similar wall shape are plotted together with proposed wall shape with using same Spitzer resistivity profile for all three cases.
High-$n$ growth rates are close to those with self-similar wall at $b=1.08a$, whereas low-$n$ rates are similar to the self-similar wall at $b=1.2a$. 

\subsubsection{Density profile vs. uniform density}~~\\

The influence of non-uniform pedestal density profile on the stability of edge modes has been studied. 
Density pedestal has driven the edge localized modes  more unstable, as overall growth rate of all modes increases higher than the uniform density case (Fig.~8). The growth rates of $n=2-4$ modes are nearly same for both density cases but more different for $n=5-10$ modes.
The higher the toroidal mode number is, the stronger is the influence of density pedestal on growth rate. Here, level of uniform density is kept same as the value of density profile at magnetic axis, therefore the normalizing Alfven time scale ($\tau_A=6.627 \times 10^{-7}s$) is same for both density cases.  

\subsubsection{Mode structures}~~\\

The detailed structure of modes $n=3,10$ are shown in contour plots of Fig.~9a-d for self-similar wall located at $b=1.2a$, and also these are shown in Fig.~10a-d for the proposed wall geometry of CFETR. The perturbed pressure and radial component of magnetic field quantities are plotted in 2D (R-Z) plane for modes $n=3,10$. All unstable modes have radial structure only localized at the edge pedestal region which predicts them to be of peeling-ballooning types. The location of all modes
is close to the inside of separatrix which is indicated by black lines of poloidal flux contour. The positions and shapes of mode structure  of these two different wall configurations remain same in (Figs.~$9-10$). The spatial structure of the $n=10$ mode is more radially localized than that of the $n=3$ mode. 

\subsubsection{Convergence test}~~\\

A thorough convergence has been checked for radial and poloidal grid numbers, time step ($\Delta t$) and polynomial degree of finite element basis used in NIMROD calculation. The growth rates of modes $n=3,10$ remain 
almost same for poloidal grid number range $150-240$ (Fig.~10a) and radial grid number range $60-96$ (Fig.~10b). From time step $\Delta t = 5 \times 10^{-9}s$ to $\Delta t = 5 \times 10^{-8}s$ (Fig.~10c) the variation in growth rate remains within $1\%$. Although there is moderate difference in growth between polynomial degree $4$ and $5$ for mode $n=10$, but polynomial degrees $5$ and $6$ have almost same growth rates (Fig.~10d). 
These results show a good numerical convergence in our calculation.  

\section{Summary and Discussions}
In summary, our analysis on the linear stability of CFETR  baseline equilibrium, finds the excitation of edge localized modes at the pedestal region but no global modes are found to be dominantly unstable. Two different resisitivity models have been employed in the calculation, 
namely the hyperbolic tangent profile and the Spitzer resistivity profile. The growth rates of $n=1-10$ have been separately calculated and compared for these resisitivity models. In the ideal MHD model using hyperbolic tangent resistivity profile, 
modes $n=1-10$ are found to be unstable with edge localized mode structure. 

The effect of conducting wall position on the stability of ideal MHD modes have been evaluated. A noticeable difference is found between the results from two resistivity profiles. In Spitzer resistivity profile case, all modes become stabilized before wall position $b=1.04a$ but for hyperbolic tangent profile, all modes 
remain unstable even if the wall is placed at plasma boundary. 

On basis of our analysis, the baseline scenario of CFETR equilibrium is not expected to dominantly unstable to global ideal MHD modes. This might help to avoid disruption event caused by such ideal MHD instabilities. But, due to steep pedestal gradient and peaked edge current, this scenario
can be susceptible to the medium to large size ELMs. 

This present calculation draws an overall picture of unstable linear ideal MHD modes with perfectly conducting wall in the CFETR baseline scenario, which are dominantly edge-localized modes in nature. To achieve long duration 
steady state operation maintaining fixed $\beta_N$, efficient methods need to be investigated for controlling ELMs. Further characteristics of ELMs need to be determined from nonlinear simulation.  In addition, the effect of toroidal flow on ELMs in this CFETR baseline scenario is planned to be examined as another potential element for changing ELM characteristics.     

\ack
The research was supported by the National Magnetic Confinement Fusion Program of China under Grant Nos. 2014GB124002, 2015GB101004, and the 100 Talent Program of the Chinese Academy of Sciences (CAS). Author D.~B. is partially supported by CAS 
President International Fellowship Initiative (PIFI), the China Postdoctoral Science Foundation
Grant No. 2016M592054 and the Anhui Provincial Natural
Science Foundation Grant No. 1708085QA22. Author P.~
Zhu also acknowledges the supports from U.S. DOE grant
Nos. DE-FG02-86ER53218 and DE-FC02-08ER54975. We
are grateful for the support from the NIMROD team. The
numerical calcul­ations in this paper have been done on the
super computing system in the Supercomputing Center of
University of Science and Technology of China. This research
used resources of the National Energy Research Scientific
Computing Center, a DOE Office of Science User Facility supported by the Office of Science of the U.S. Department
of Energy under Contract No. DE-AC02-05CH11231.
\section{References}
\bibliographystyle{iopart-num}
\bibliography{cfetr_pst}
\newpage
\begin{figure}[htbp]
\subfloat[]{\includegraphics[width=7.5cm]{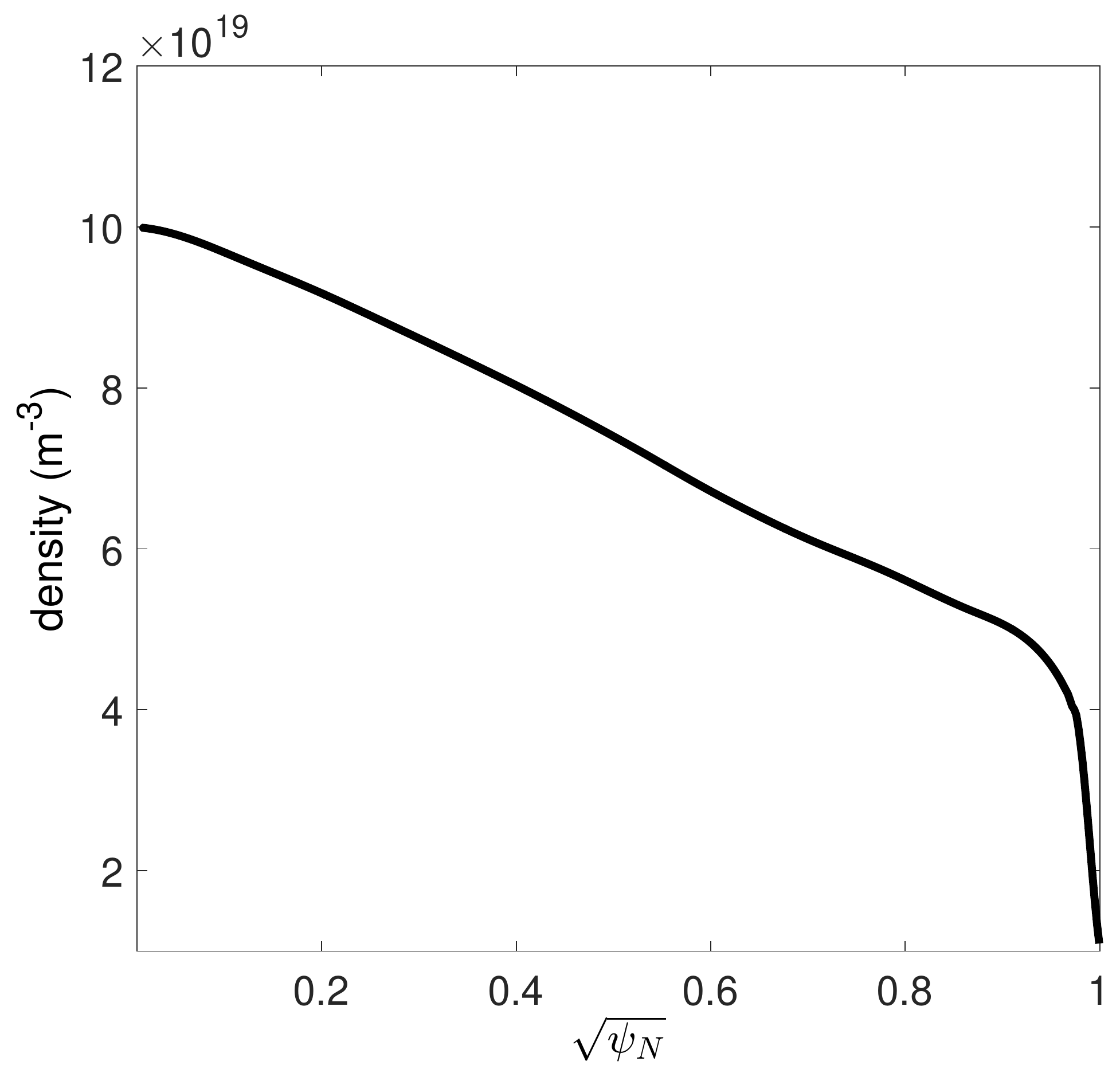}}
~~\subfloat[]{\includegraphics[width=7.5cm]{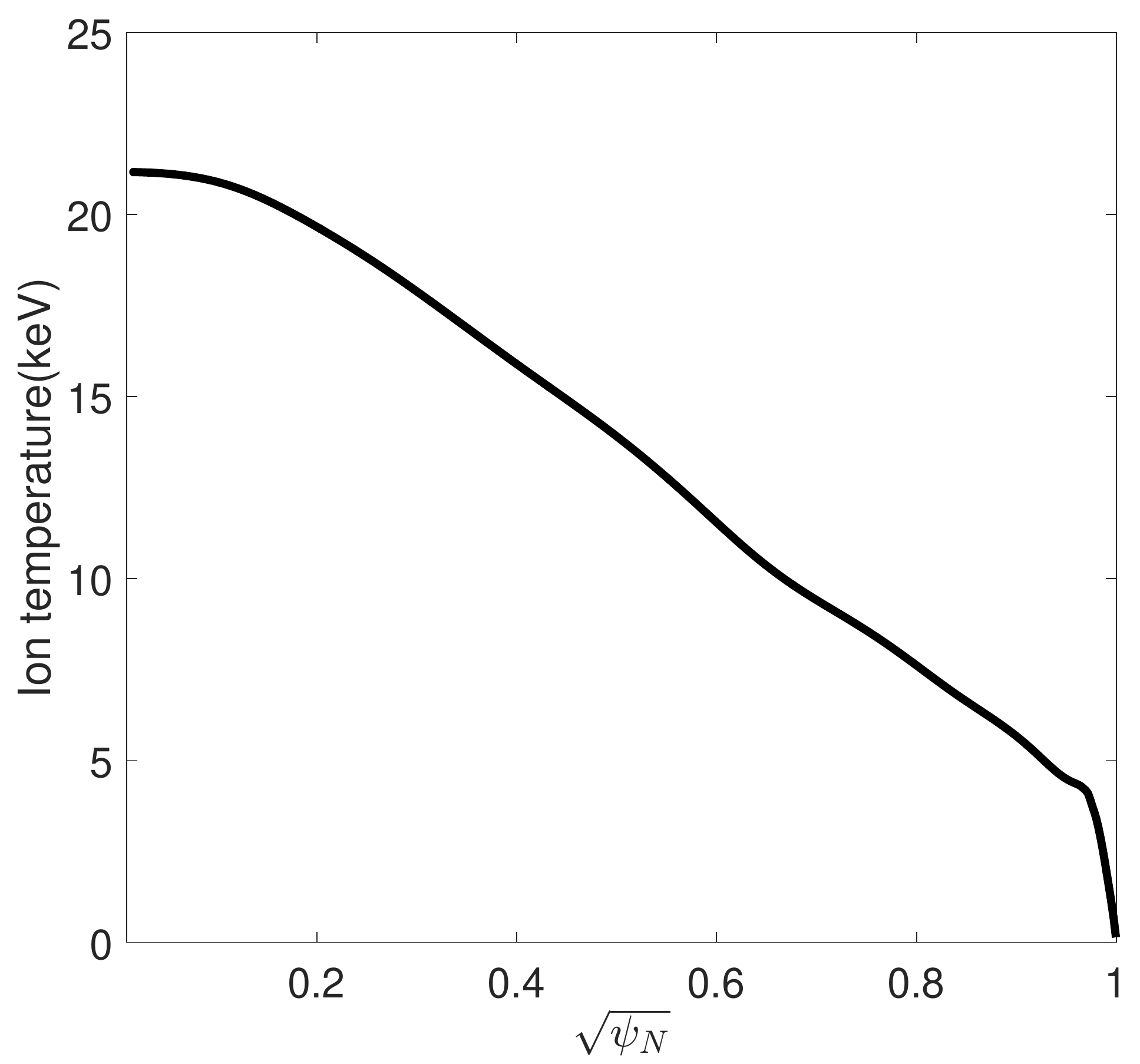}}\\
~~~~~~~~~~~~~~~~\\
~~~~~~~~~~~~~~~~\\
\subfloat[]{\includegraphics[width=7.5cm]{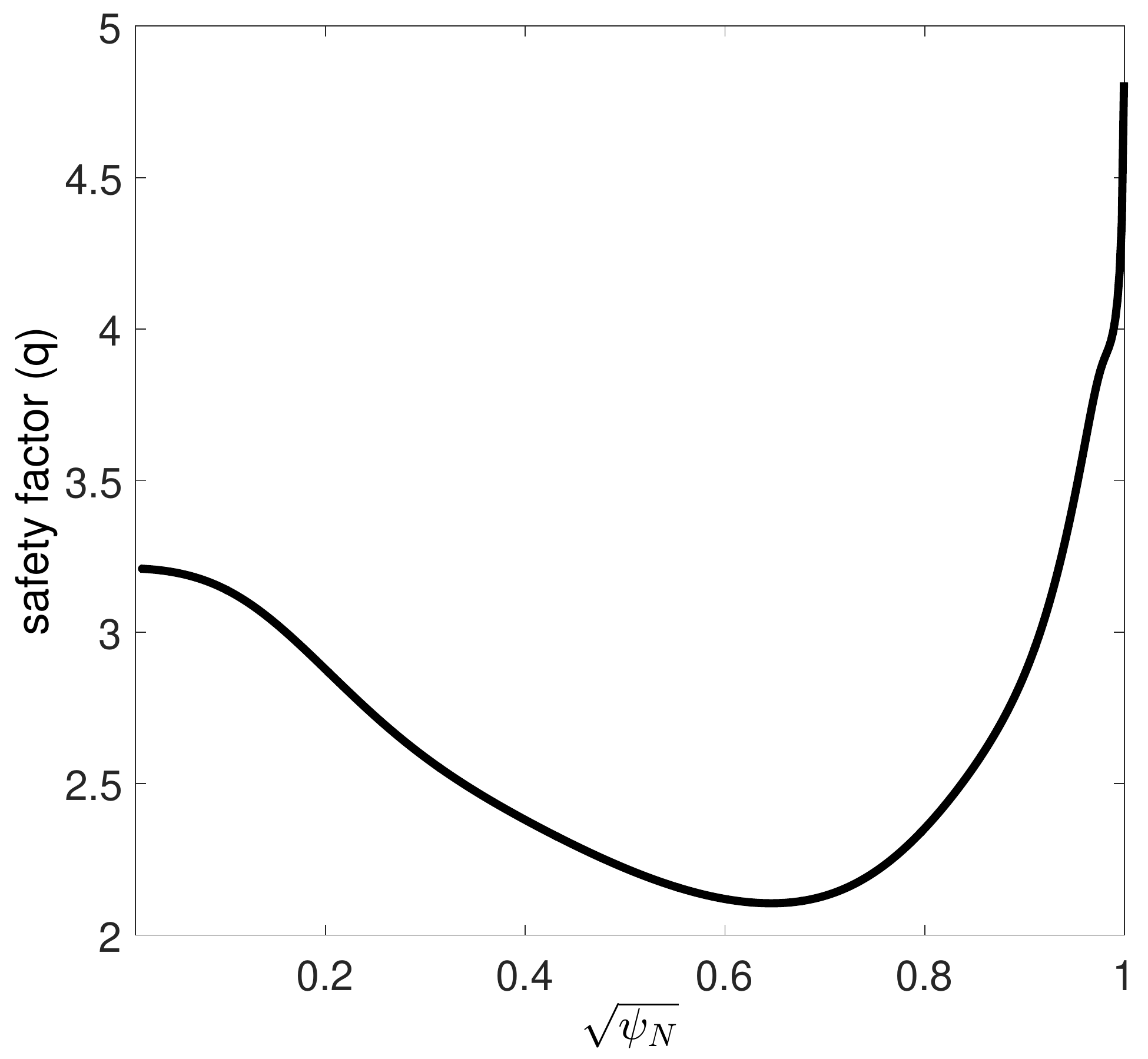}}
~~\subfloat[]{\includegraphics[width=7.9cm]{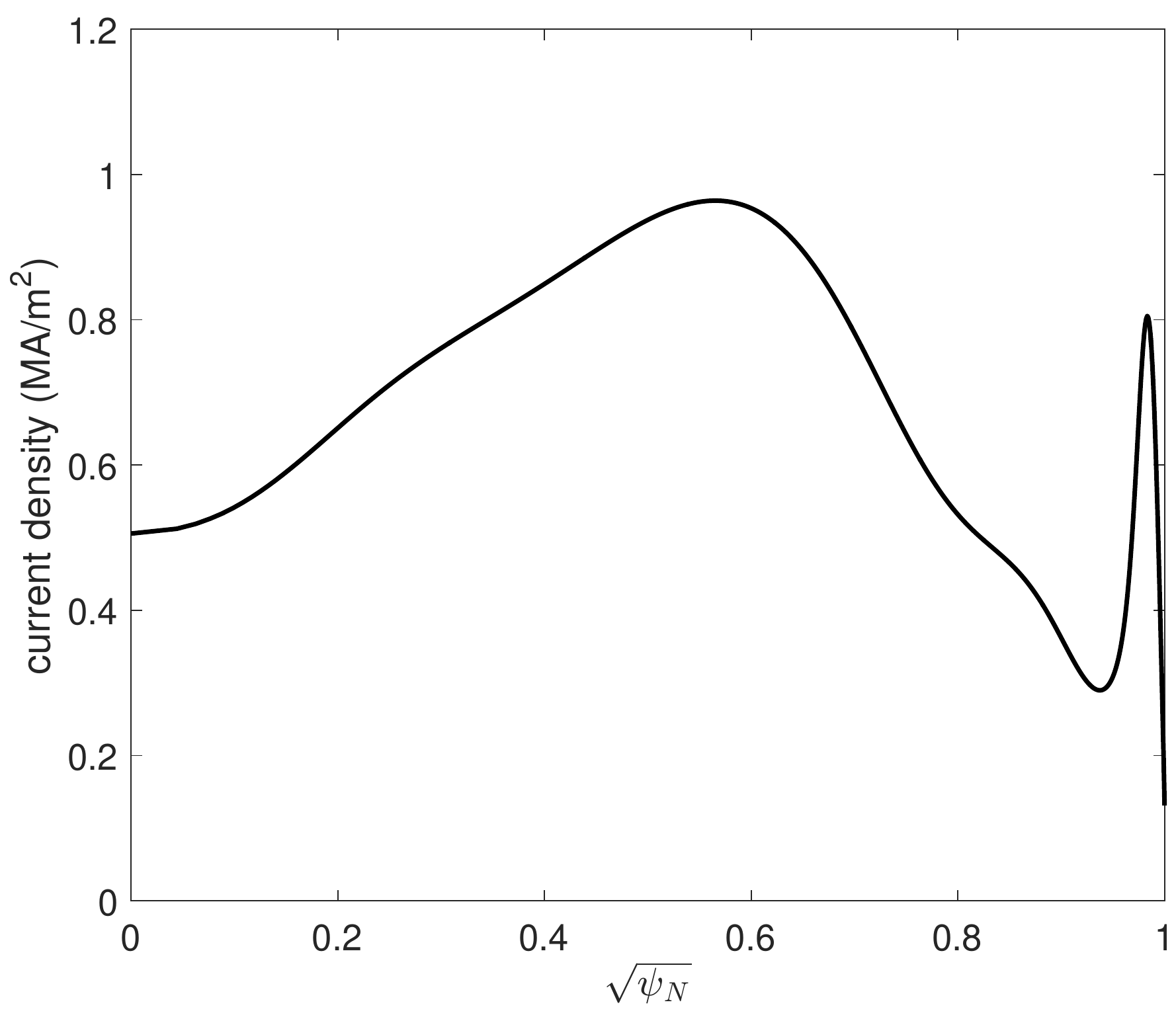}}
\caption{\label{equil} Radial profiles of electron density~(a), ion 
temperature~(b), safety factor~(c) and current density~(d) for CFETR  baseline equilibrium are drawn. $\psi_{N}$ is the normalized poloidal flux function.
Both density and temperature have high pedestal region at the edge and the current density has highly peaked edge part. The safety factor 
has $q_{min}>2$ and strong reversed shear region.}
\end{figure}

\newpage
\begin{figure}[htbp]
\includegraphics[height=8.5cm,width=8.5cm]{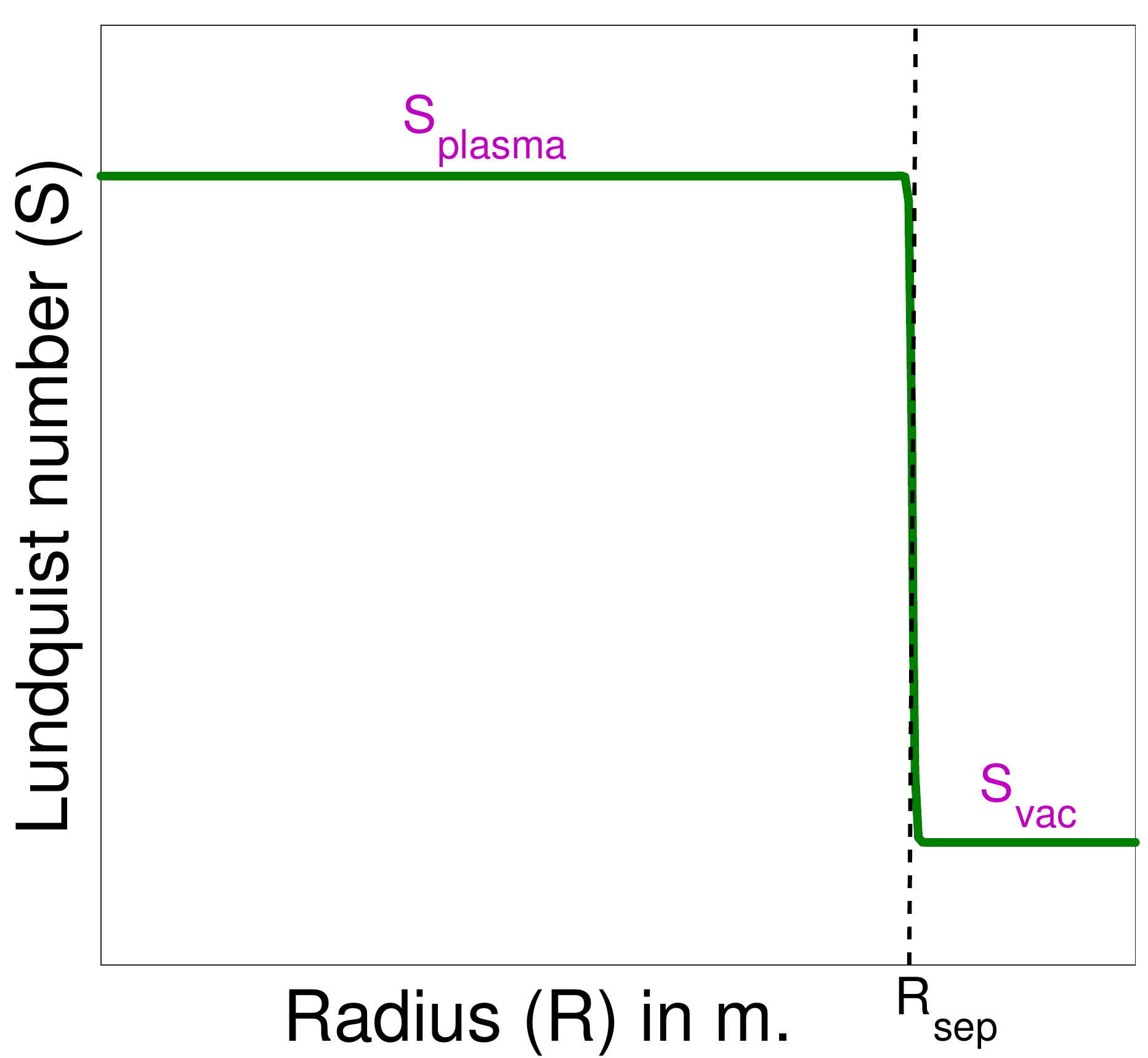}
 \caption{\label{lund} The typical profile of Lundquist number with major radius used in the simulation is drawn here. The value of $S_{plasma}$ and $S_{vac}$
are varied to define the ideal MHD limit. $R_{sep}$ is the position of separatrix. }
\end{figure}

\newpage
\begin{figure}[htbp]
\subfloat[]{\includegraphics[width=7.5cm]{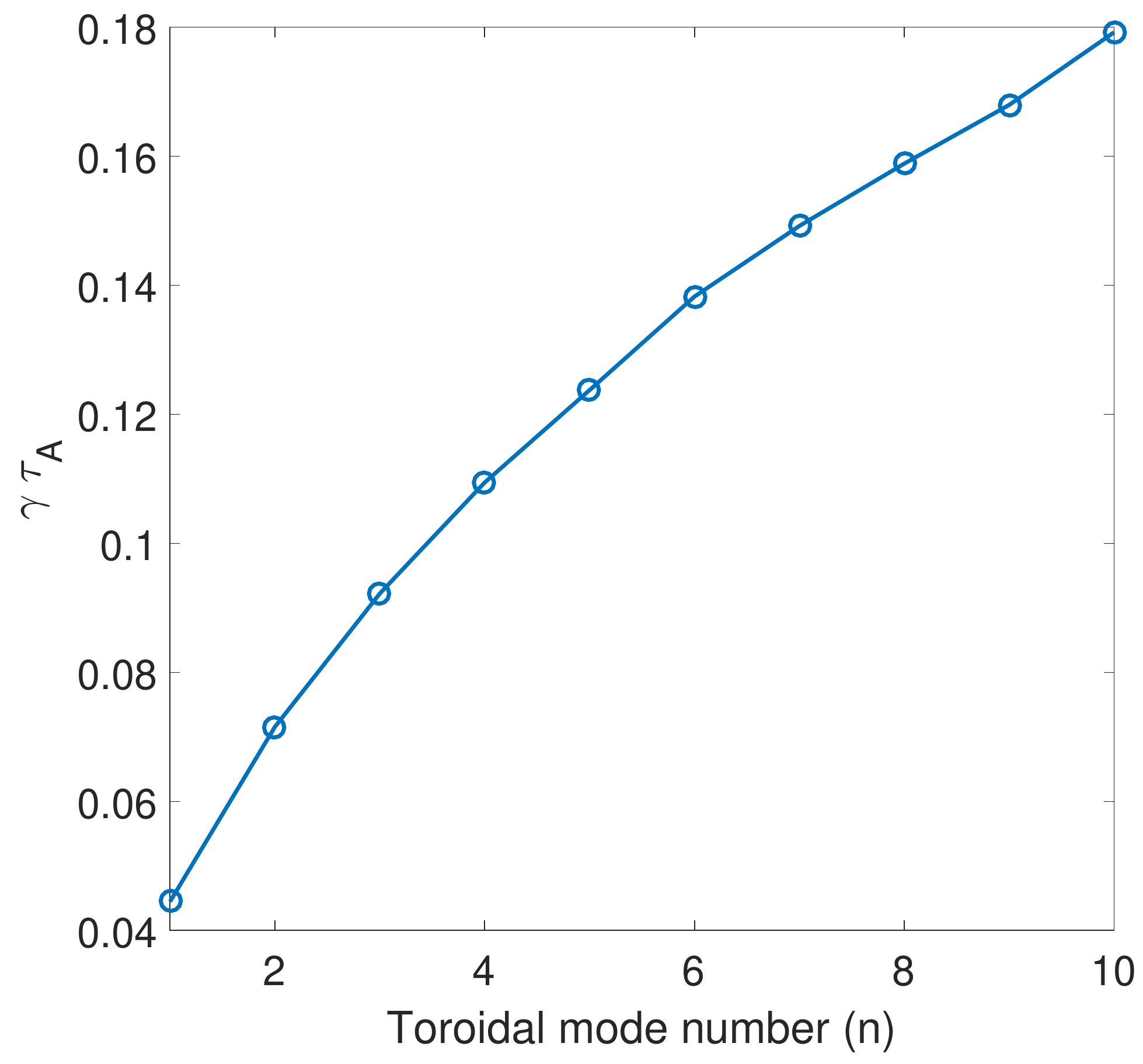}}
~~\subfloat[]{\includegraphics[width=7.5cm]{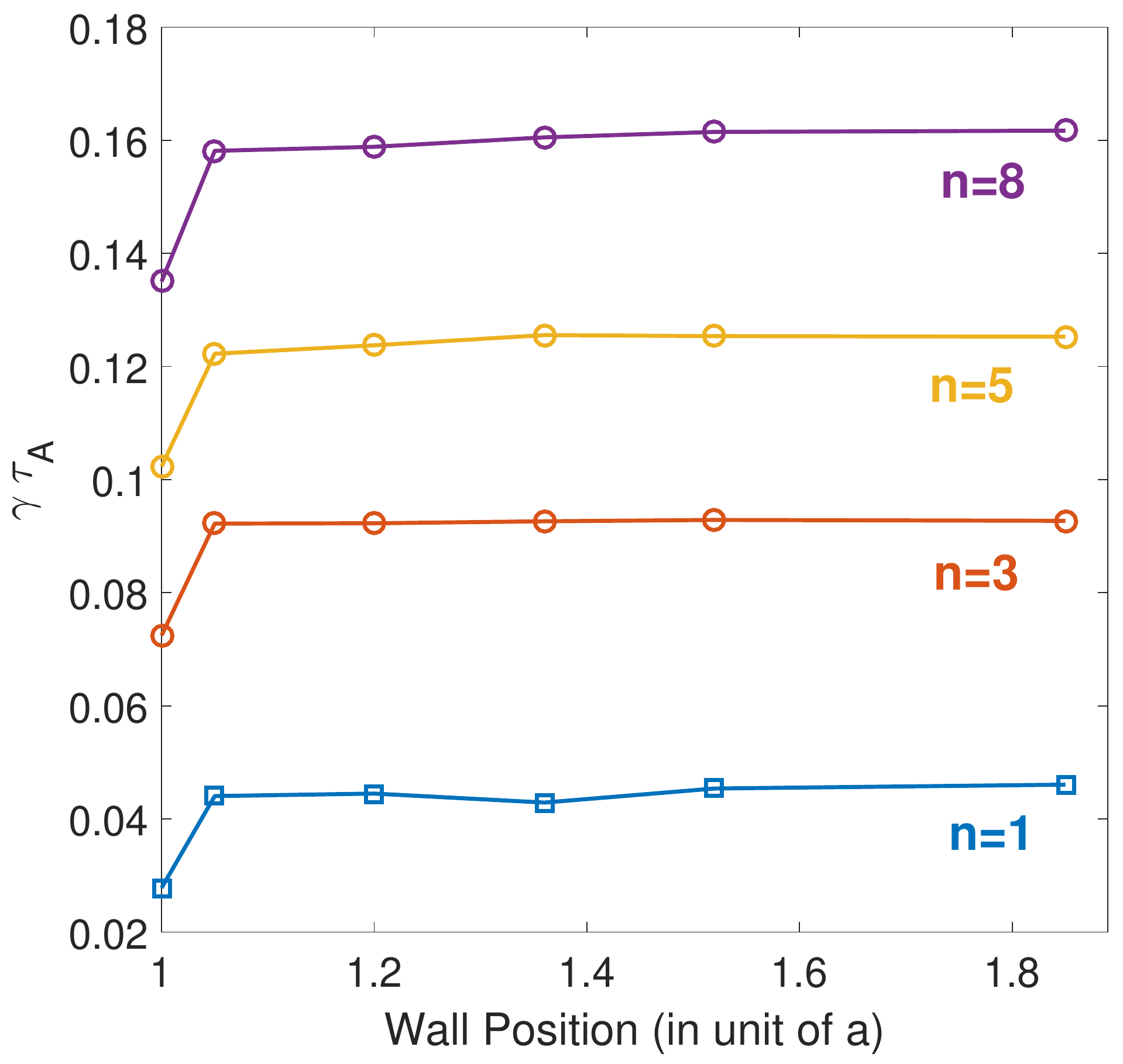}}\\
\caption{\label{grtanh} (a) Normalized growth rate of $n=1-10$ vs. toroidal mode no. $n$ in ideal MHD for wall position b=1.2a (b) Variation of growth rate of $n=1,3,5,8$
with conducting wall position. Ideal wall saturation limit is close to plasma boundary. 
}  
\end{figure}

\newpage
\begin{figure}[htbp]
\subfloat[]{\includegraphics[width=8.5cm]{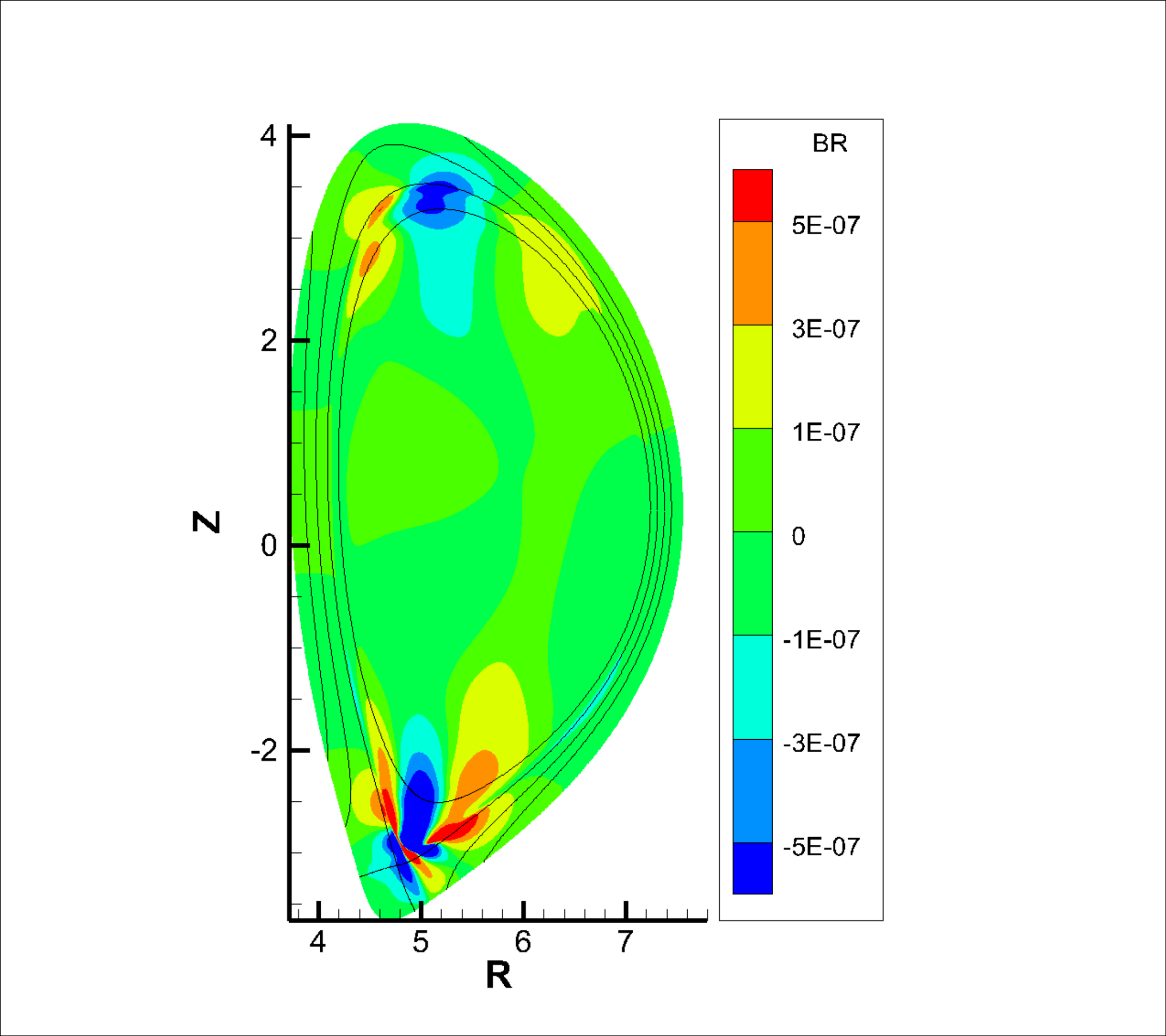}}
~~\subfloat[]{\includegraphics[width=8.5cm]{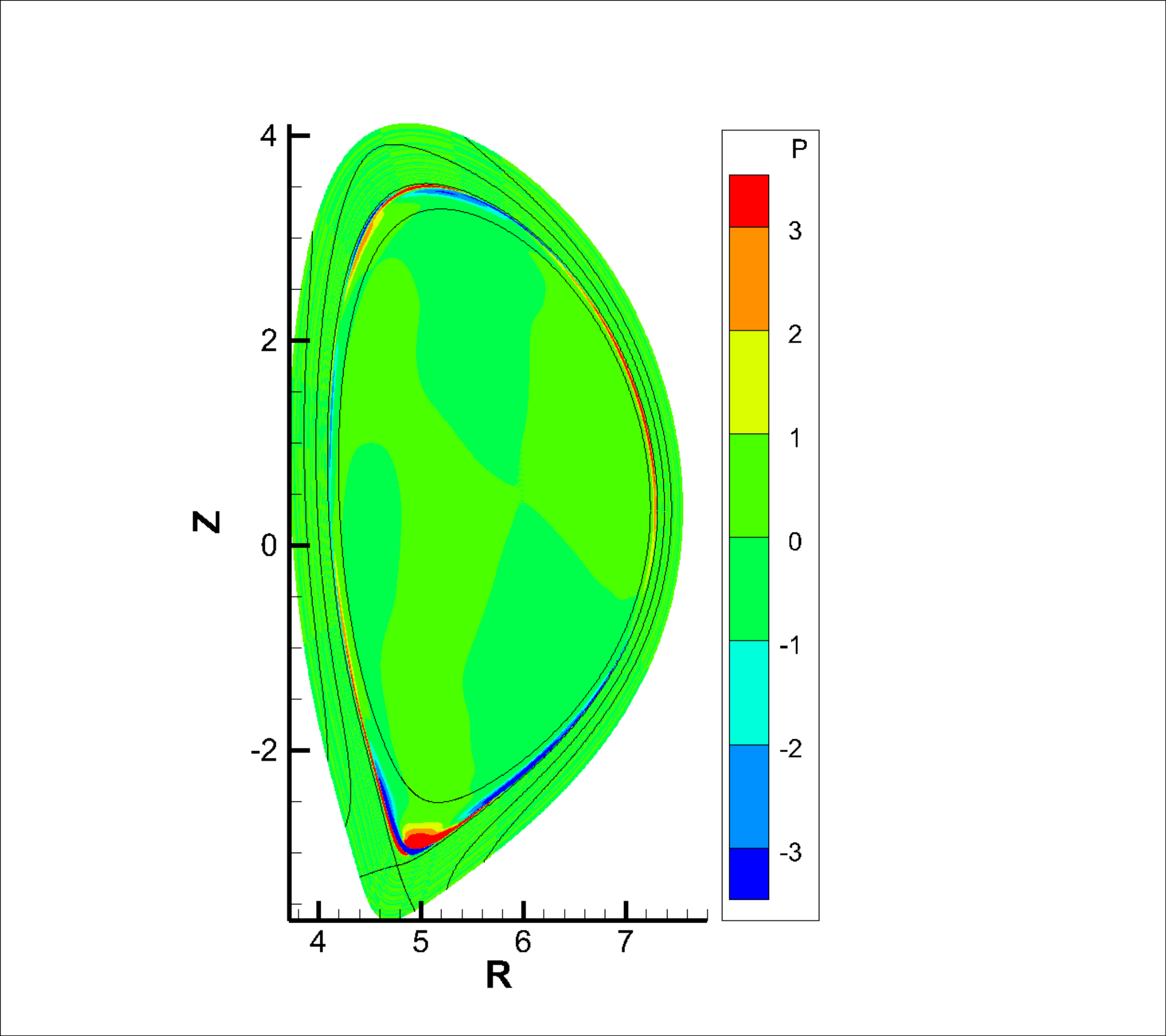}}\\
\subfloat[]{\includegraphics[width=8.5cm]{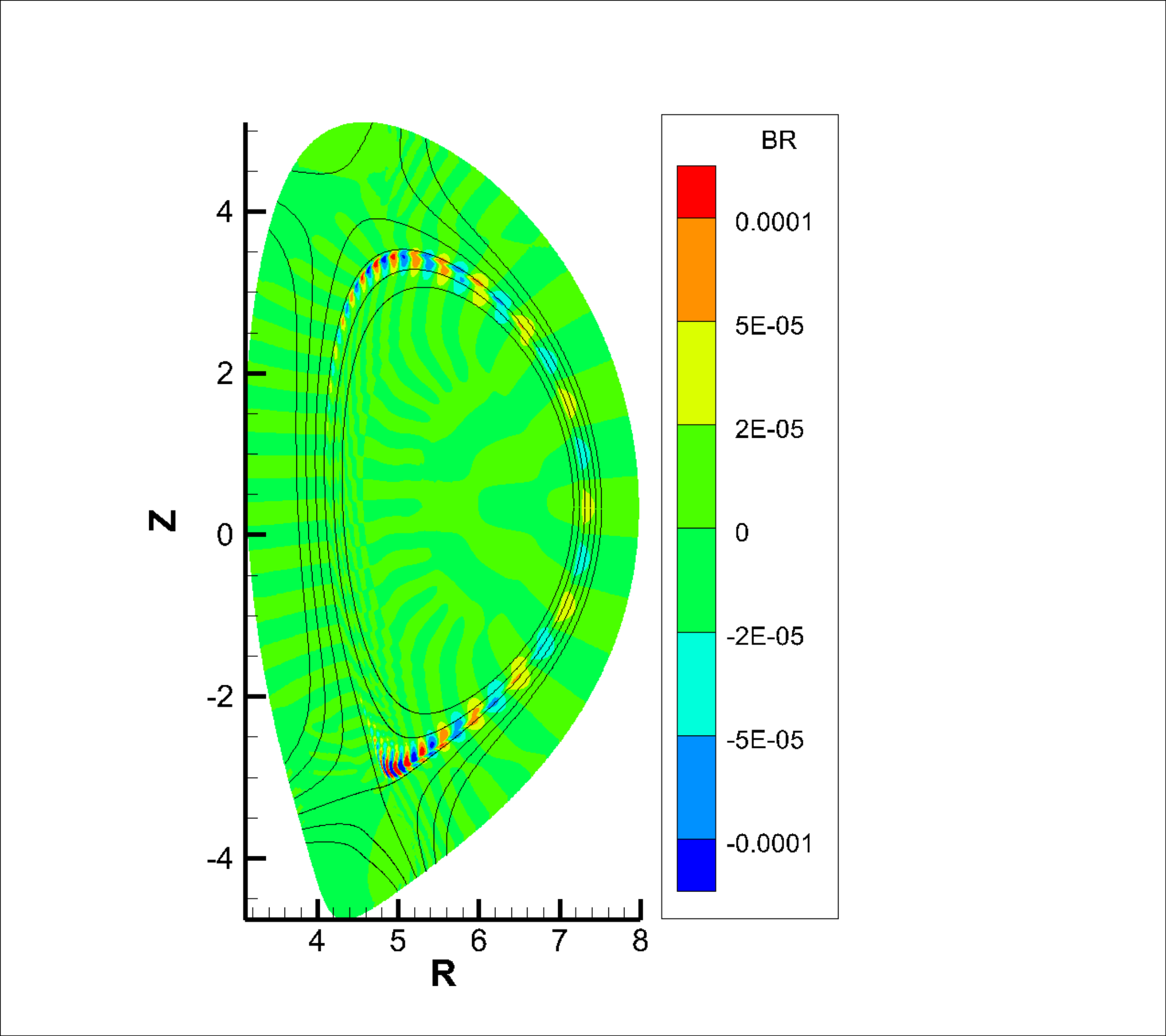}}
~~\subfloat[]{\includegraphics[width=8.5cm]{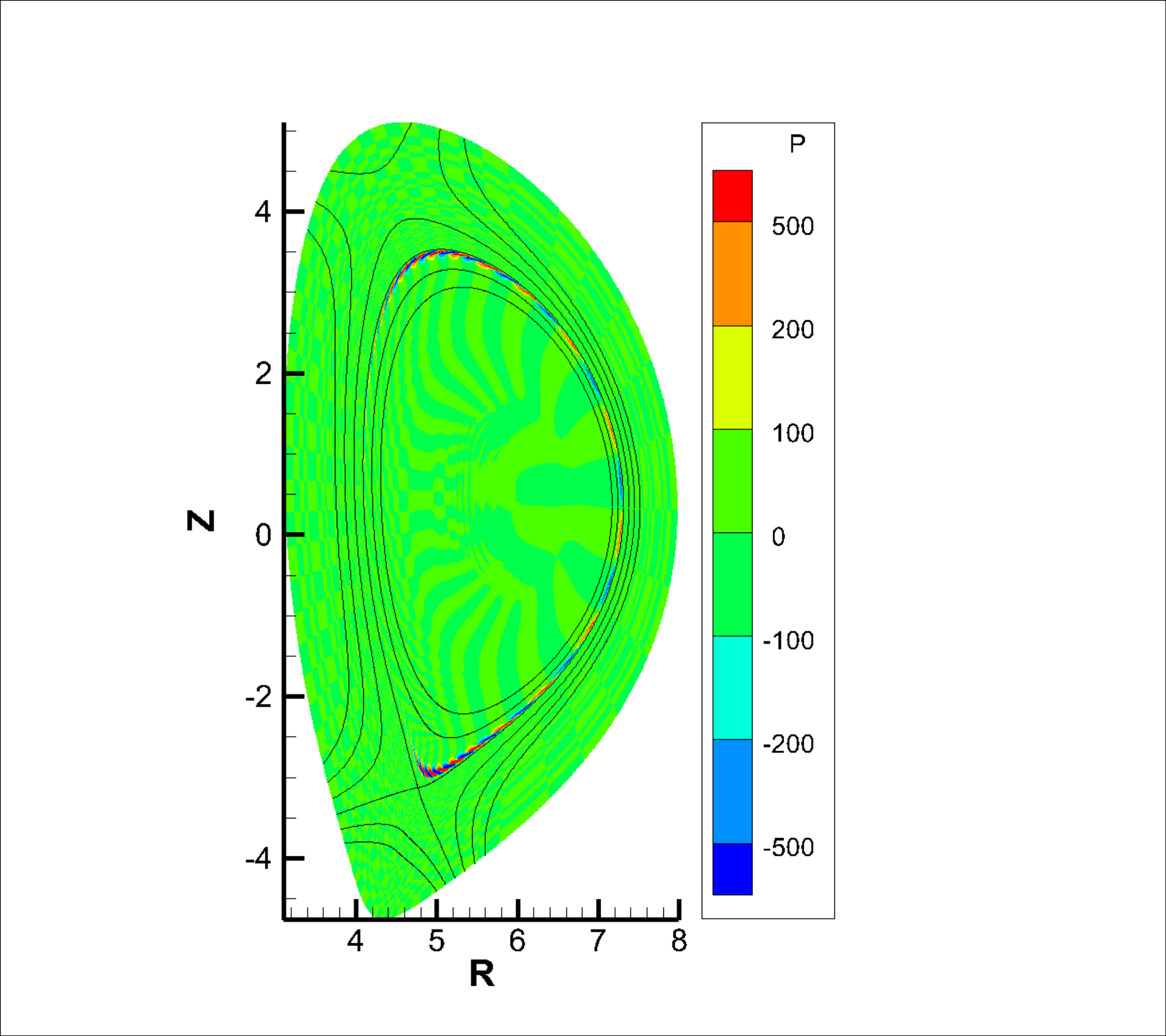}}\\
 \caption{\label{grnm} (Hyperbolic tangent profile of resisitivity) (a) Contour plot of radial component of perturbed magnetic field ($B_r$) on 2-d (R-Z) plane for n=1, (b) Perturbed pressure (P) contour for n=1, 
(c) perturbed $B_r$ contour for $n=8$, (d) perturbed P contour for $n=8$. Each mode has only edge localized structure close to separatrix (shown in black line).}
\end{figure}

\newpage
\begin{figure}[htbp]
\subfloat[]{\includegraphics[width=7.5cm]{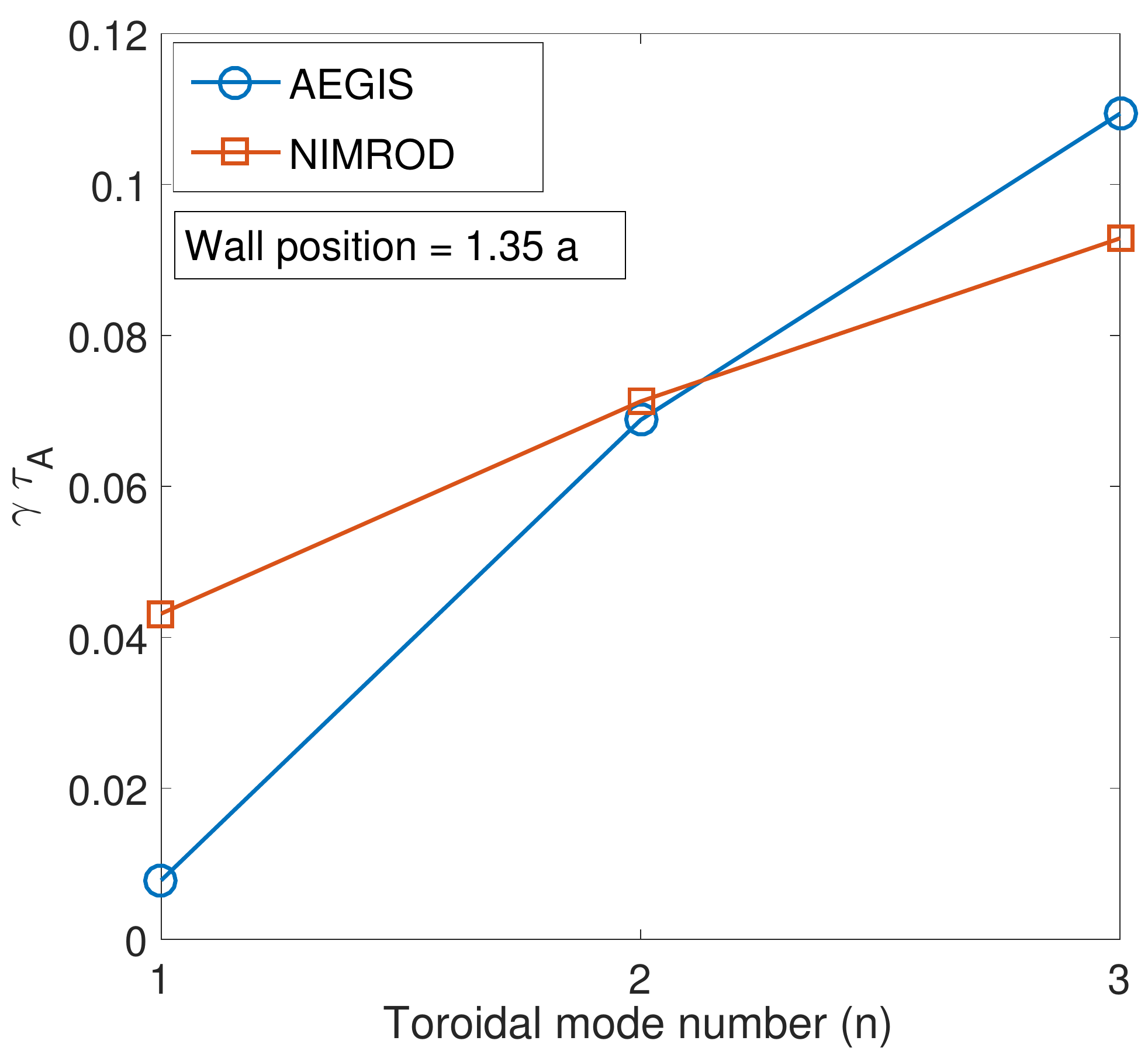}}
~~\subfloat[]{\includegraphics[width=7.5cm]{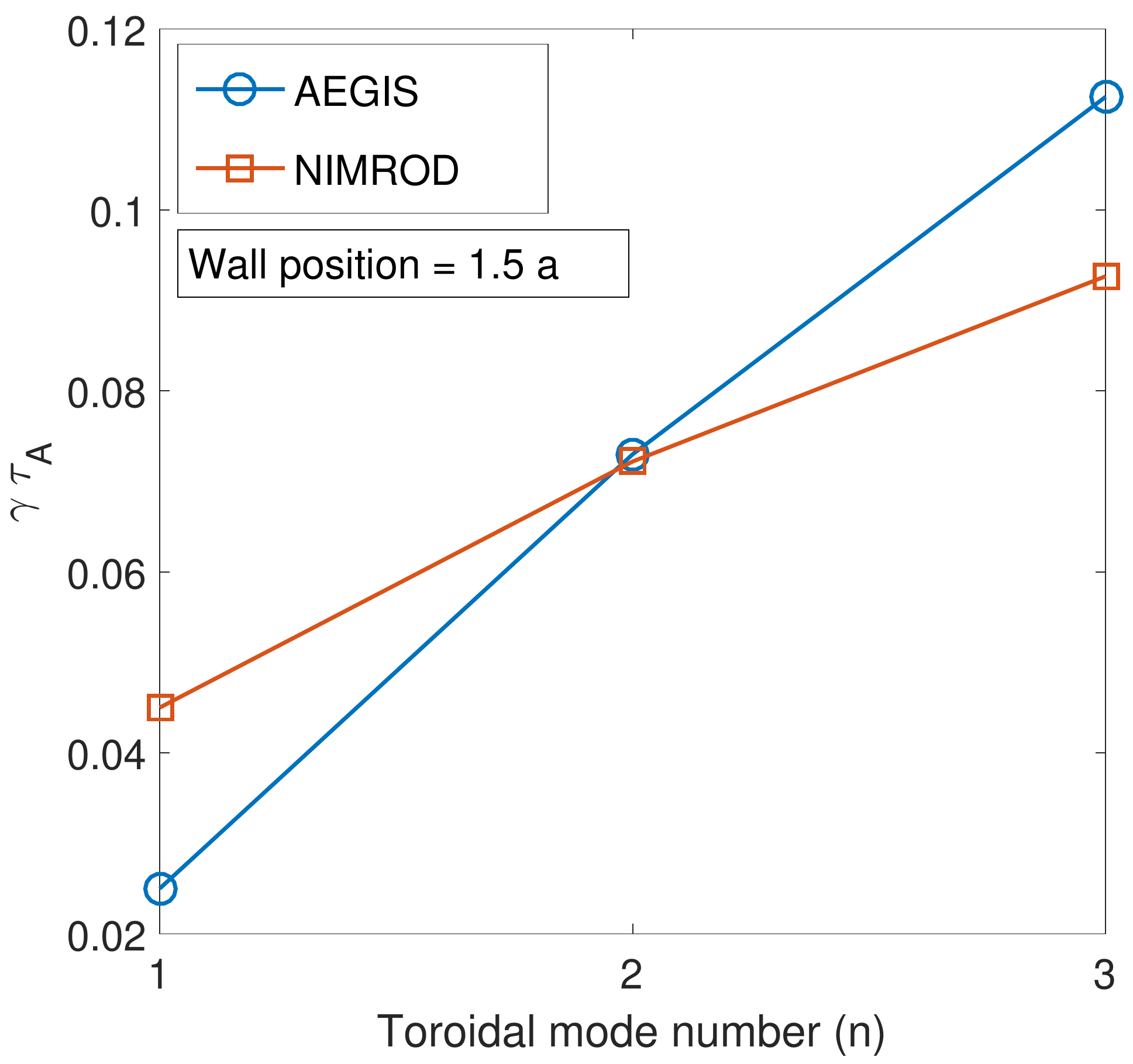}}\\
\caption{\label{contour} Growth rates of $n=1-3$ modes from NIMROD and AEGIS are plotted for conducting wall position at $1.35a$ (a) and 
 $1.5a$ (b). $n=2$ mode growth rate has good agreement between these two codes.} 
\end{figure}

\newpage
\begin{figure}[htbp]
\includegraphics[height=8.5cm,width=16.5cm]{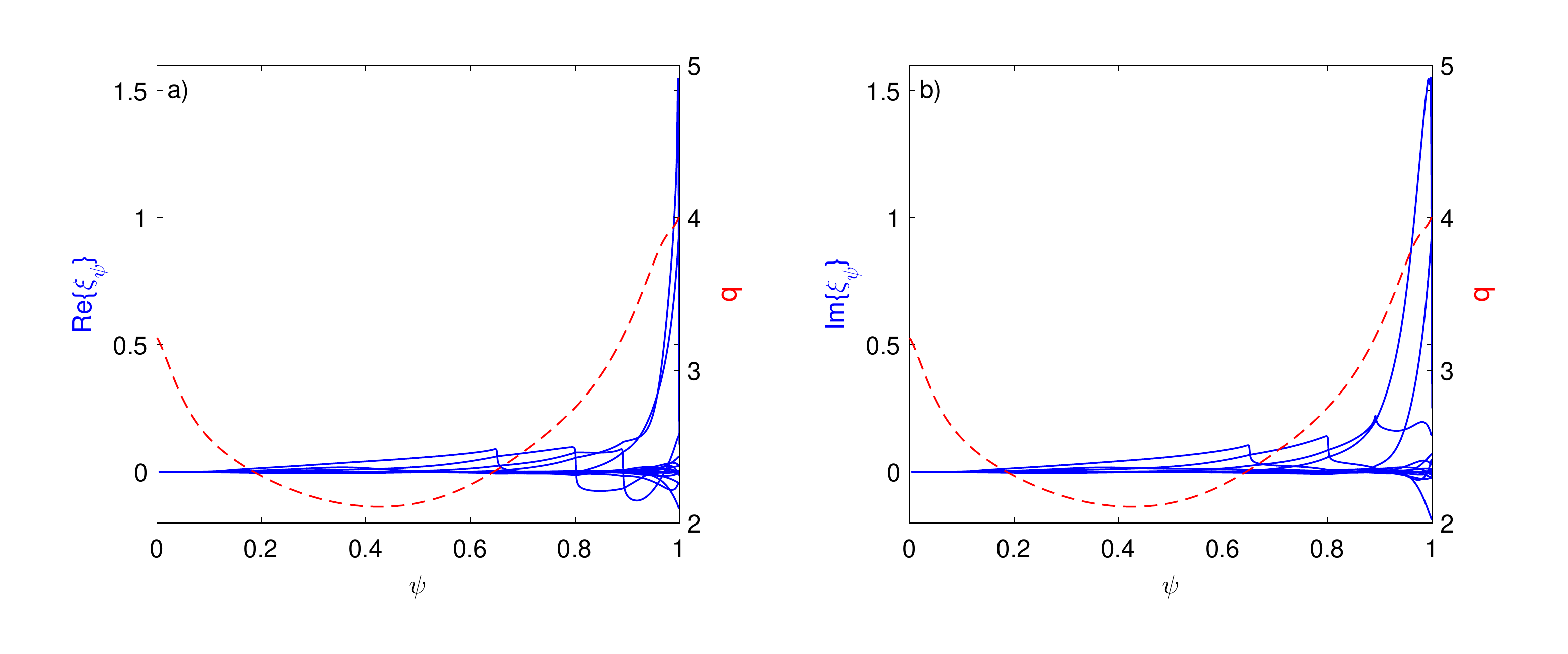}
 \caption{\label{lund} Real and imaginary component of radial displacement is shown for $n=2$ with wall position $b=1.35a$. }
\end{figure}
\newpage
\begin{figure}[htbp]
\subfloat[]{\includegraphics[width=7.5cm]{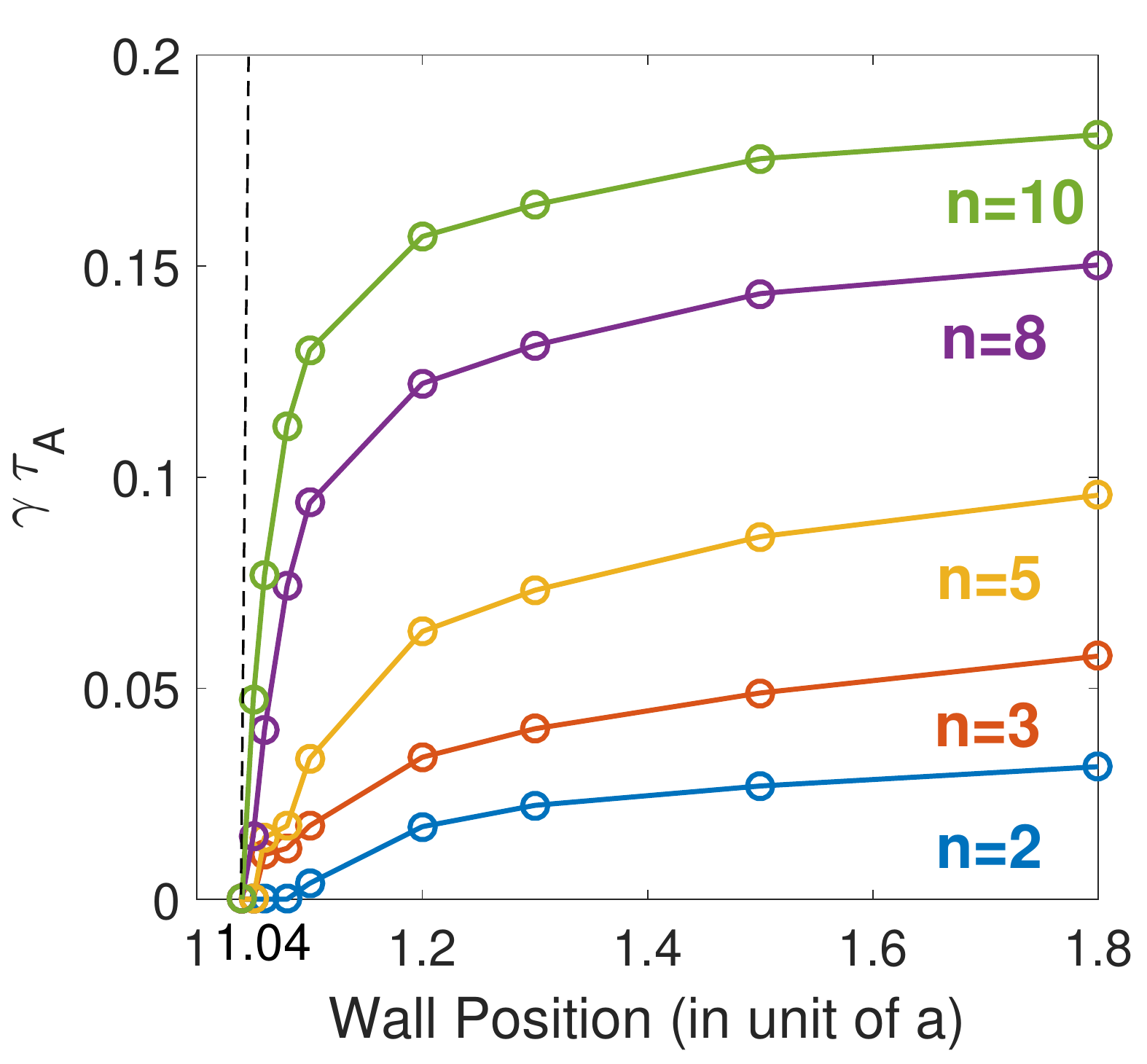}}
~~\subfloat[]{\includegraphics[width=7.5cm]{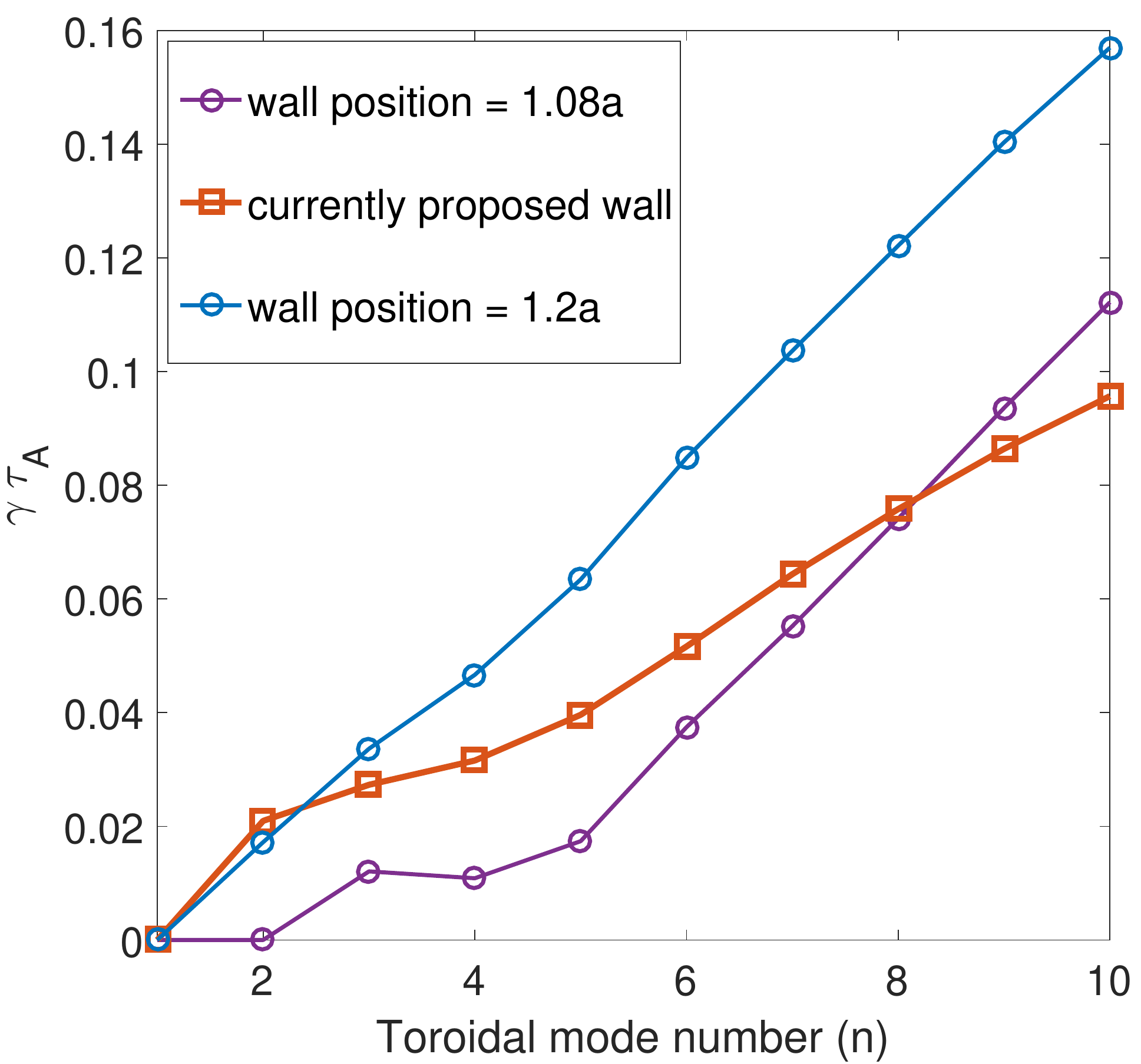}}\\
\caption{\label{wallgr} (a) Variation of normalized growth rate of $n=2,3,5,8,10$ with conducting wall position. All modes become stable at wall position $b=1.04a$. 
(b) Normalized growth rate of $n=1-10$ vs. toroidal mode no. $n$ using Spitzer resistivity profile and different shapes of wall.}  
\end{figure}
\newpage
\begin{figure}[htbp]
{\includegraphics[width=7.5cm]{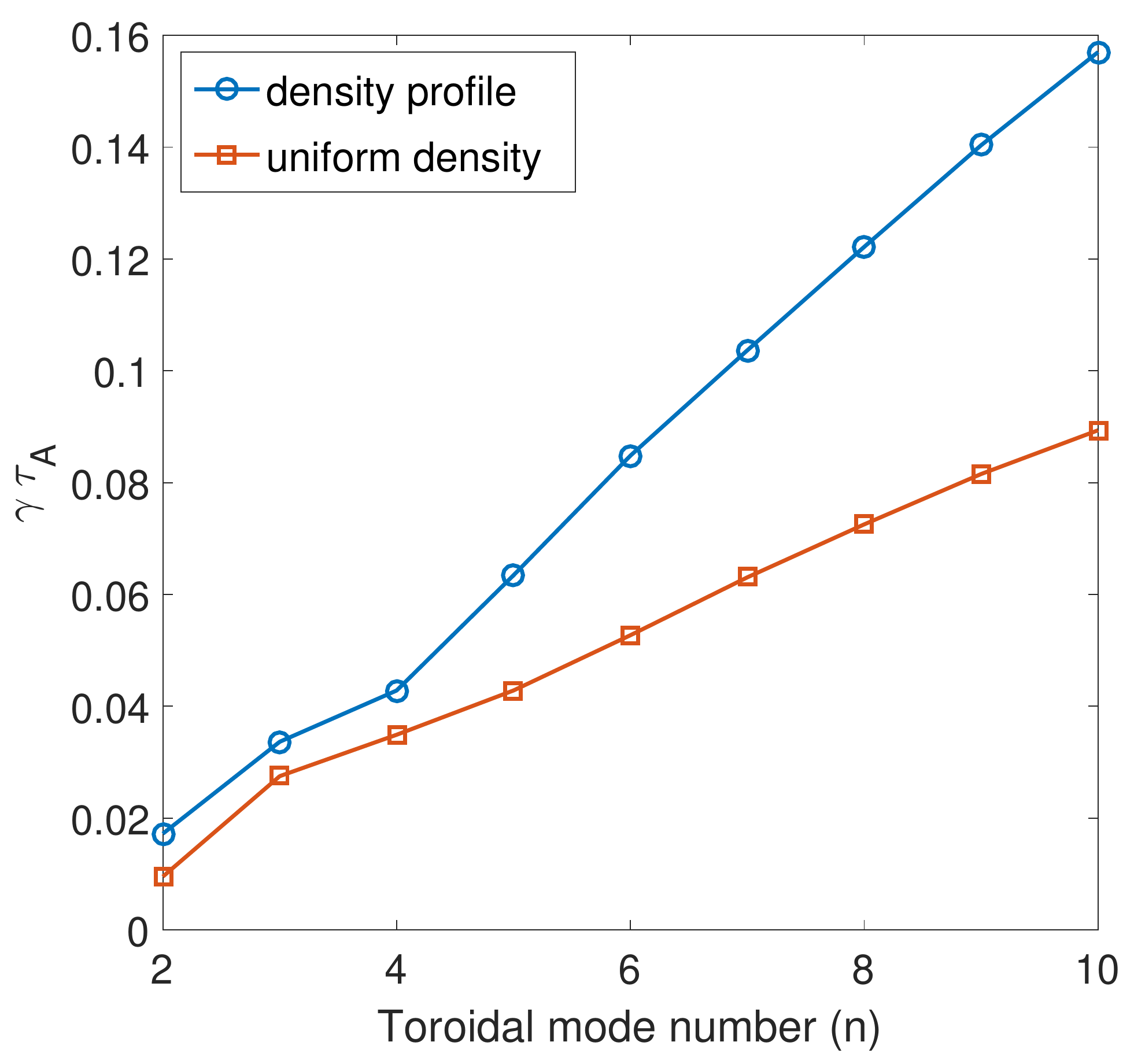}}
\caption{\label{grnm} A comparison
in growth rates for $n=2-10$ between uniform density and density profile cases. Modes $n>4$ has wide variation in growth rate.}
\end{figure}
\newpage
\begin{figure}[htbp]
\subfloat[]{\includegraphics[width=8.5cm]{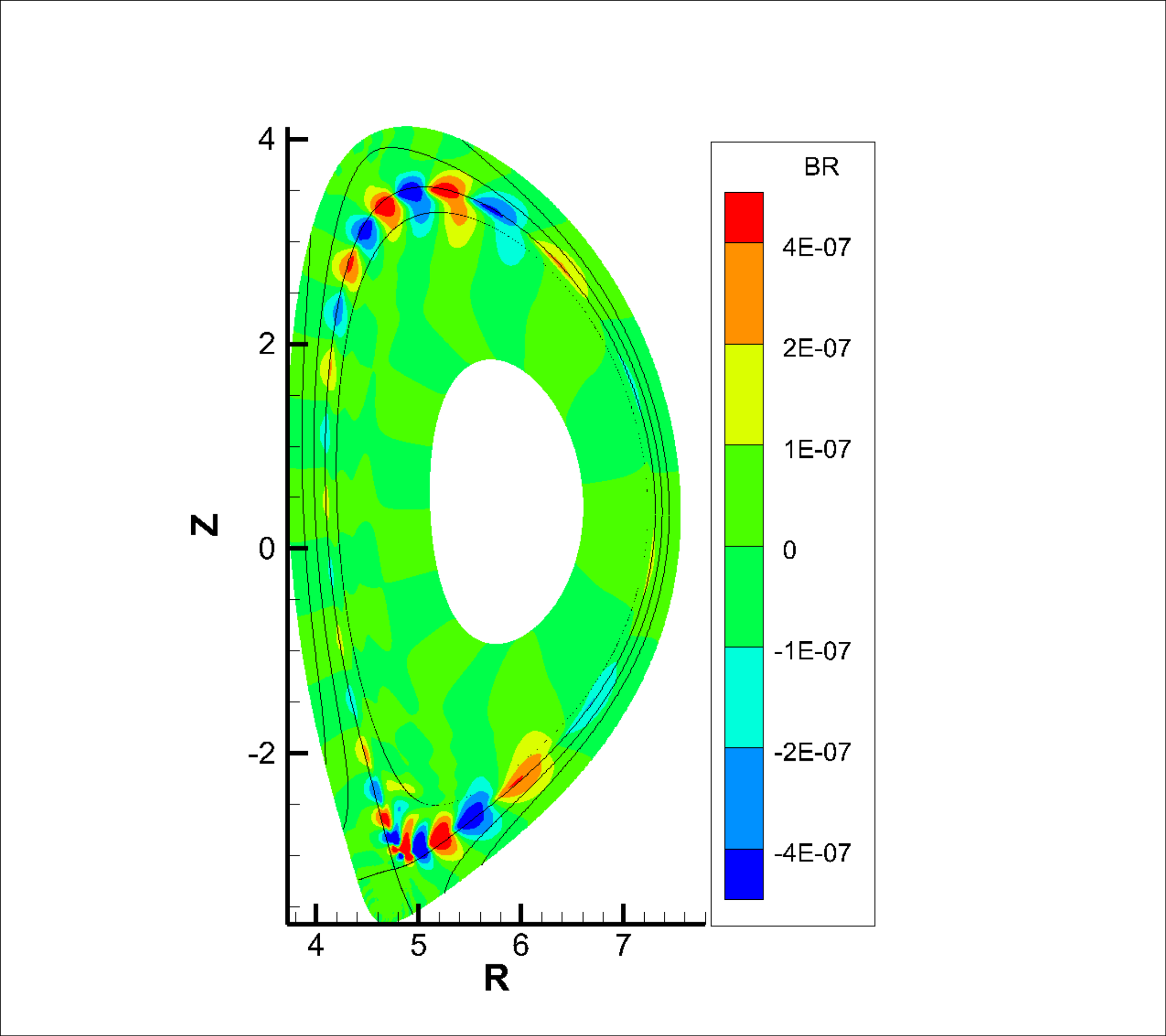}}
~~\subfloat[]{\includegraphics[width=8.5cm]{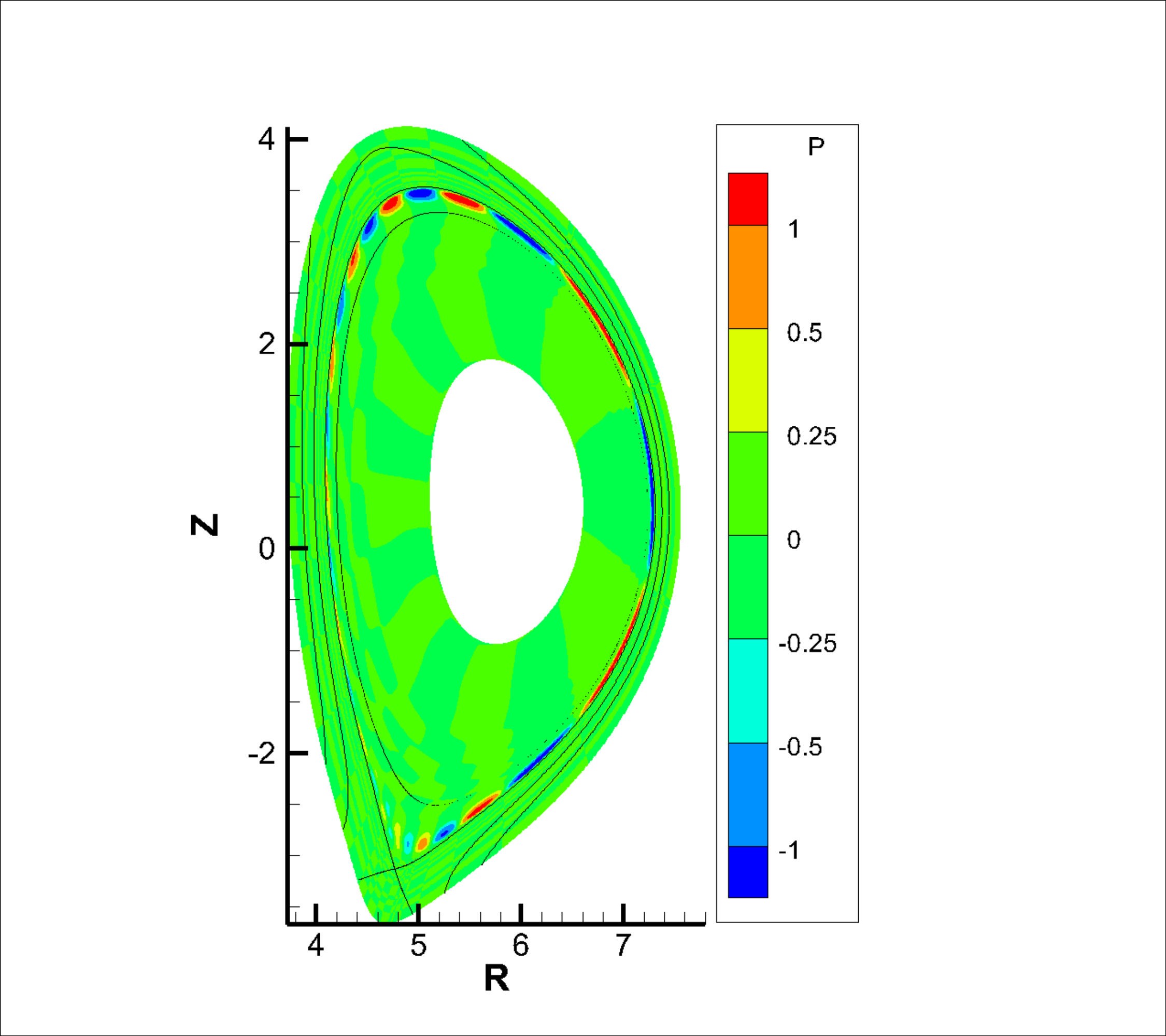}}\\
\subfloat[]{\includegraphics[width=8.5cm]{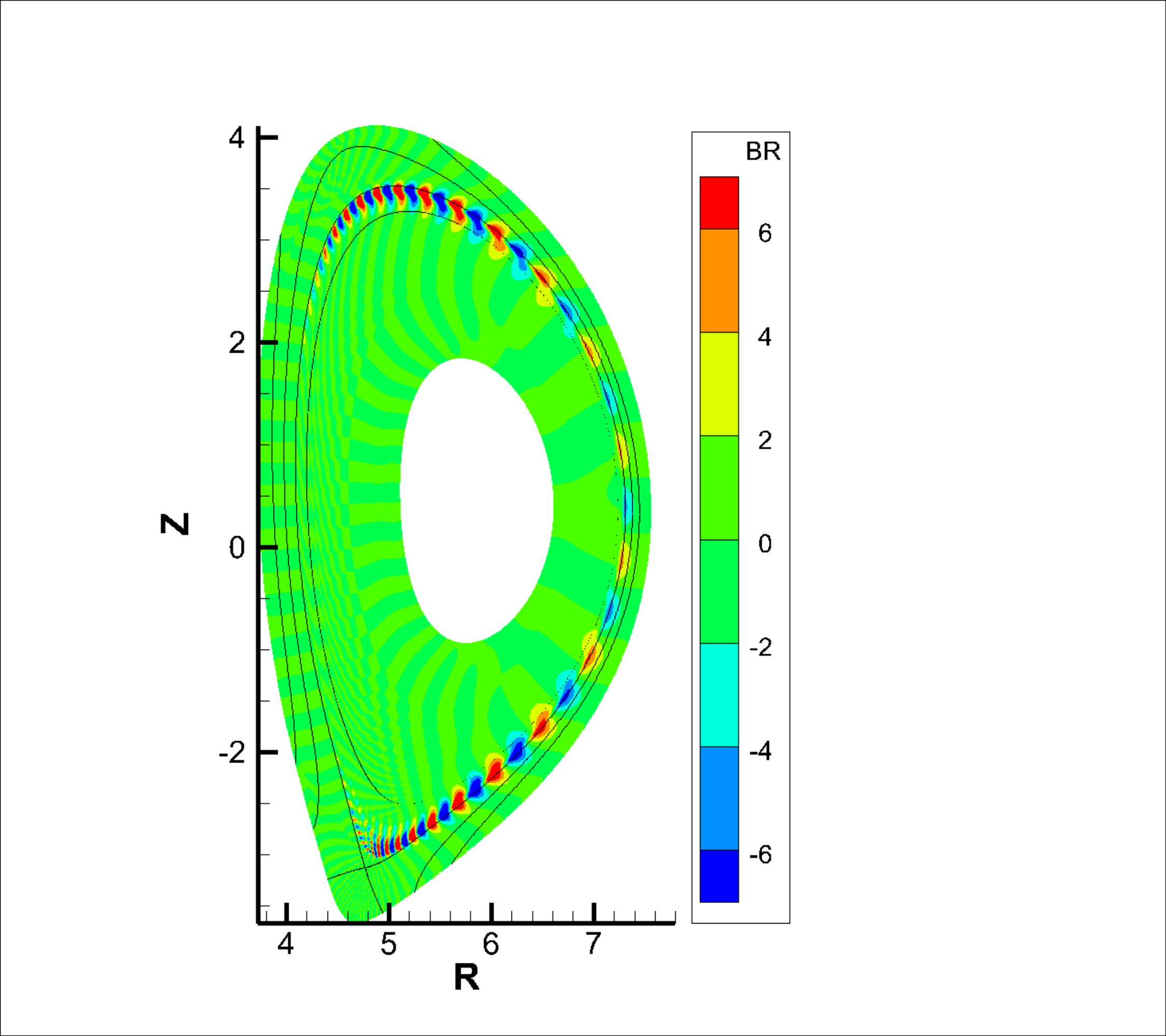}}
~~\subfloat[]{\includegraphics[width=8.5cm]{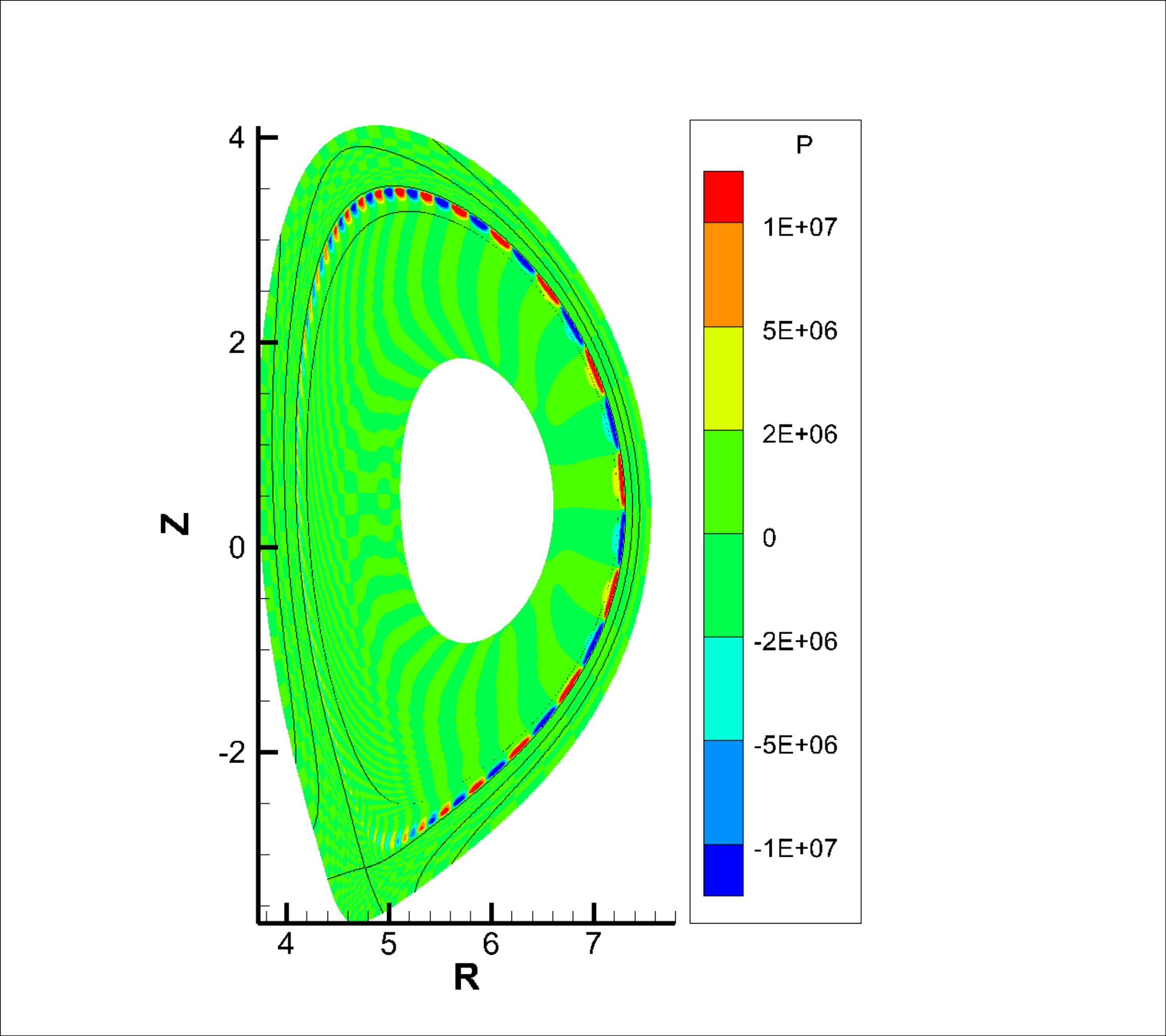}}\\
\caption{\label{contour} (Spitzer resistivity profile and self-similar wall at position $b=1.2a$) (a) Contour plot of radial component of perturbed magnetic field ($B_r$) on 2-d (R-Z) plane for n=3, (b) Perturbed pressure (P) contour for n=3, 
(c) perturbed $B_r$ contour for $n=10$, (d) perturbed P contour for $n=10$. Each mode has only edge localized structure close to separatrix (shown in black line).} 
\end{figure}
\newpage
\begin{figure}[htbp]
\subfloat[]{\includegraphics[width=8.5cm]{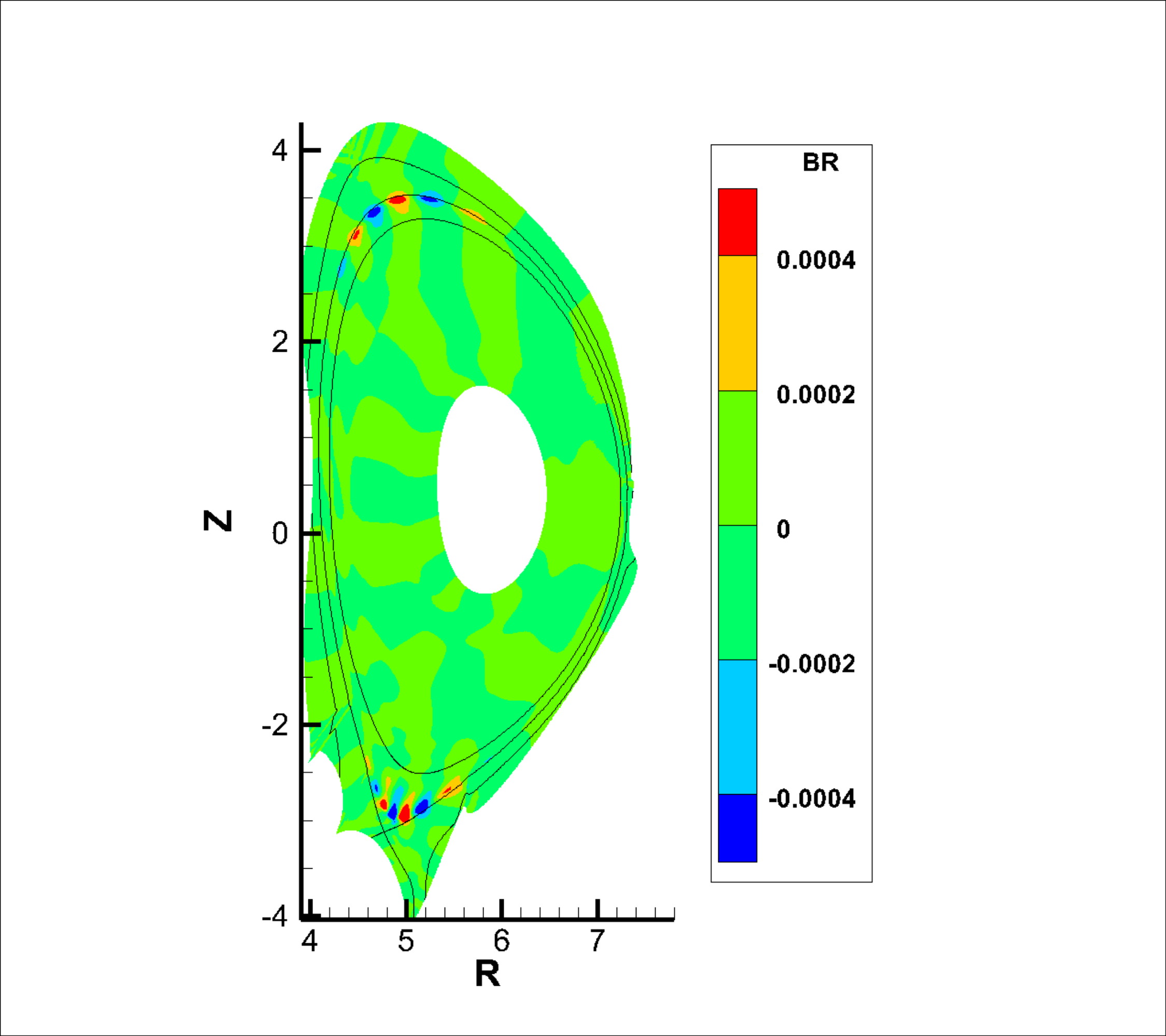}}
~~\subfloat[]{\includegraphics[width=8.5cm]{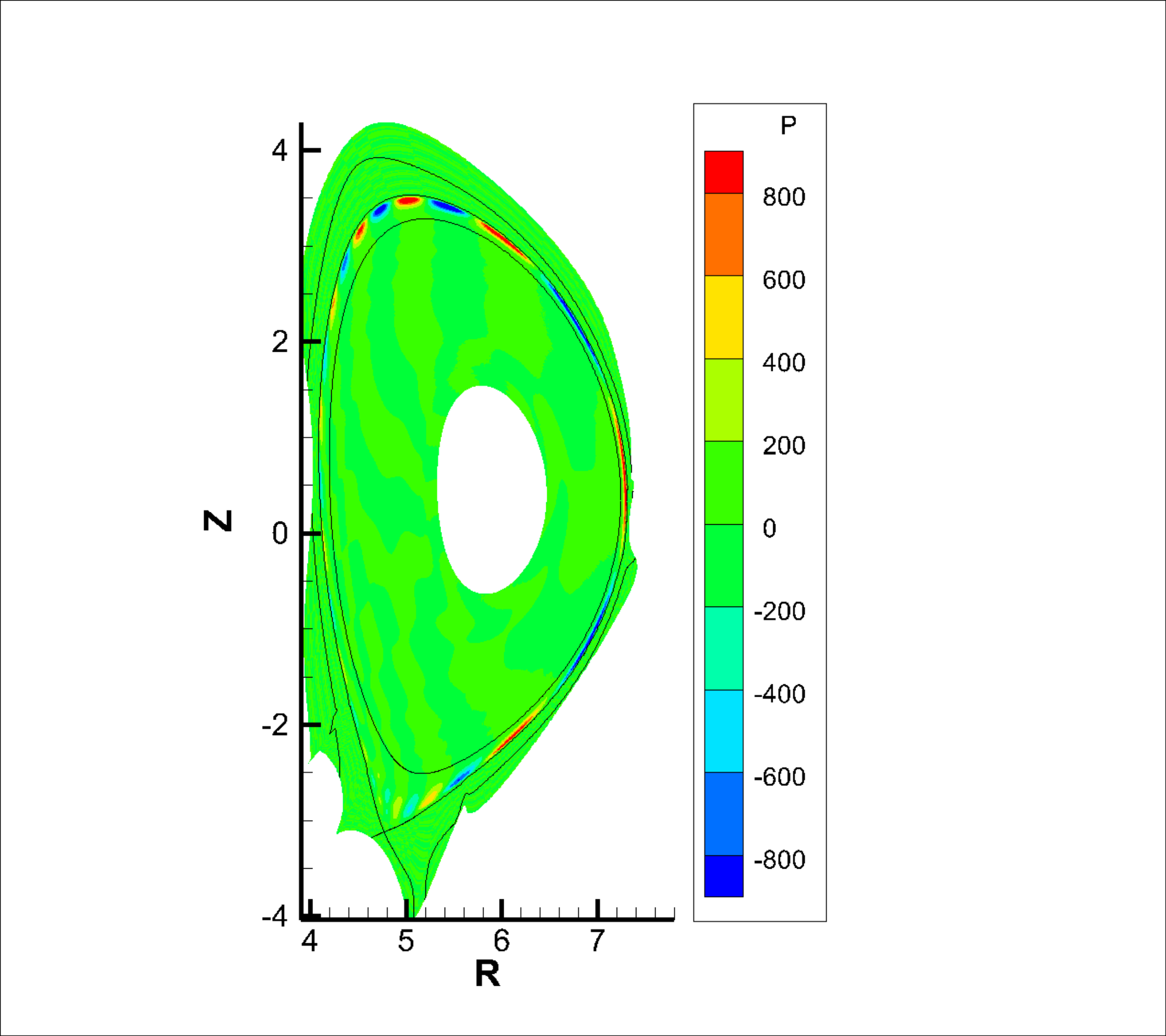}}\\
\subfloat[]{\includegraphics[width=8.5cm]{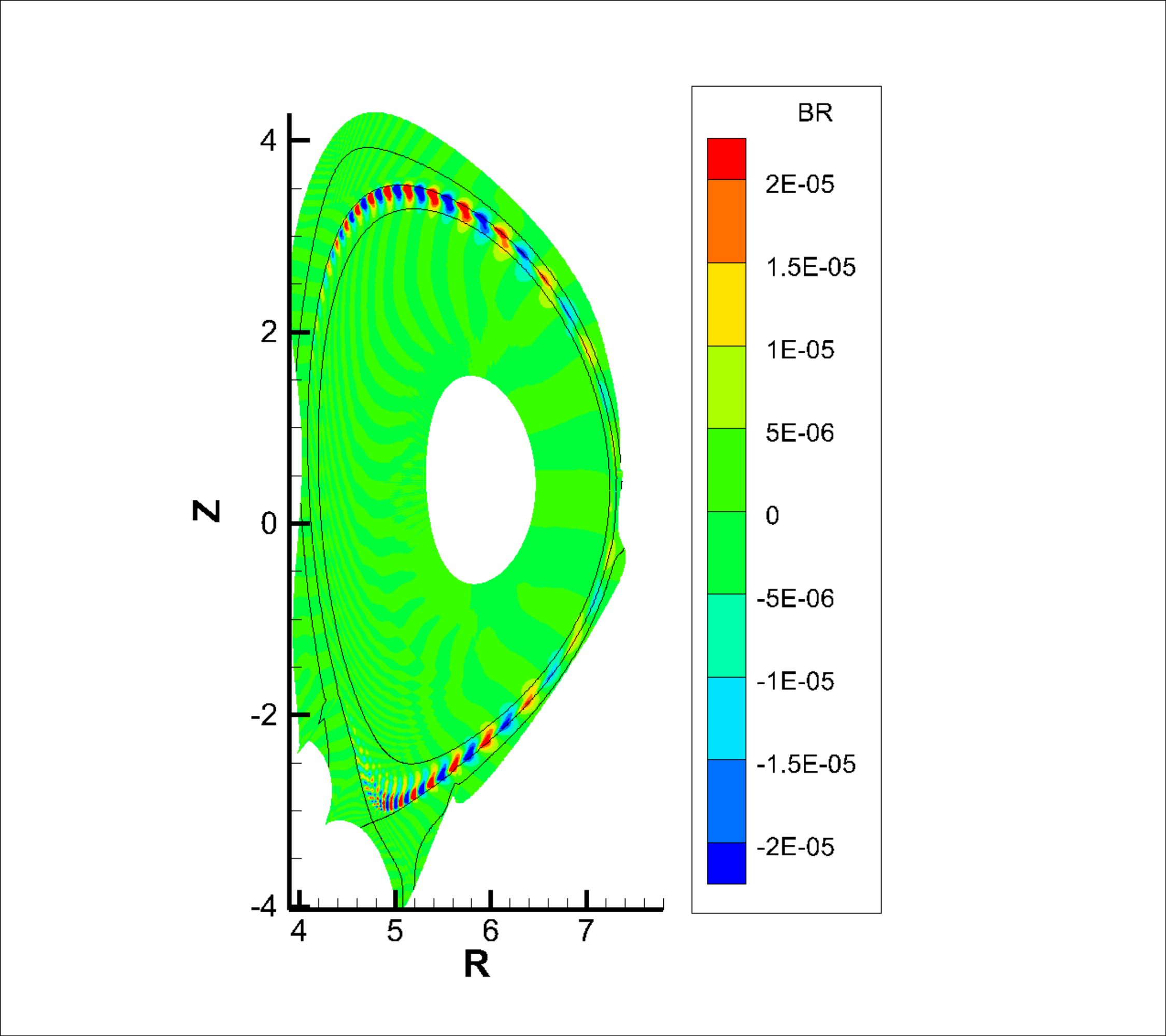}}
~~\subfloat[]{\includegraphics[width=8.5cm]{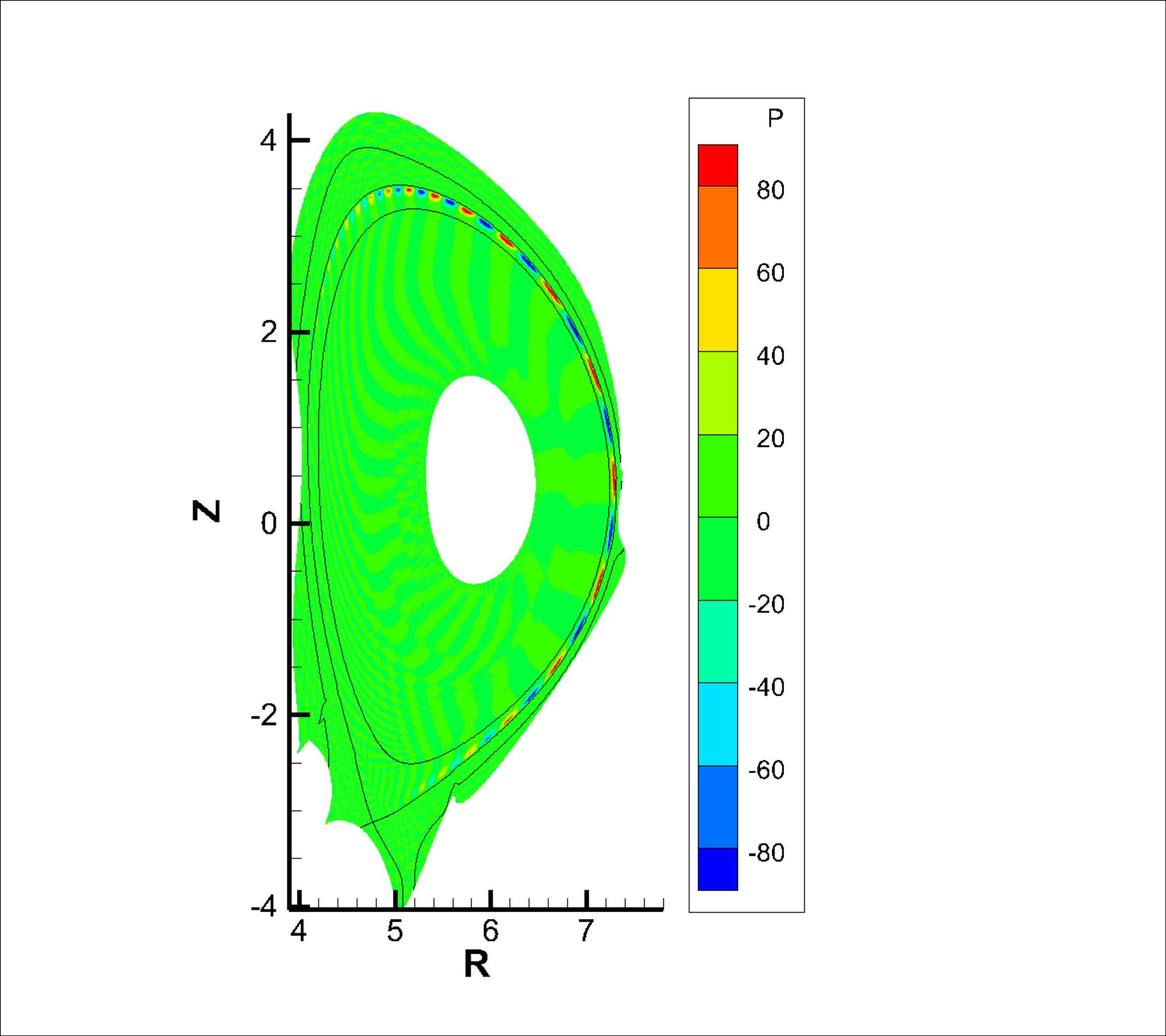}}\\
\caption{\label{contour} (Spitzer resistivity profile and present design of CFETR wall shape) (a) Contour plot of radial component of perturbed magnetic field ($B_r$) on 2-d (R-Z) plane for n=3, (b) Perturbed pressure (P) contour for n=3, 
(c) perturbed $B_r$ contour for $n=10$, (d) perturbed P contour for $n=10$. Each mode has only edge localized structure close to separatrix (shown in black line).} 
\end{figure}
\newpage
\begin{figure}[htbp]
\subfloat[]{\includegraphics[width=7.5cm]{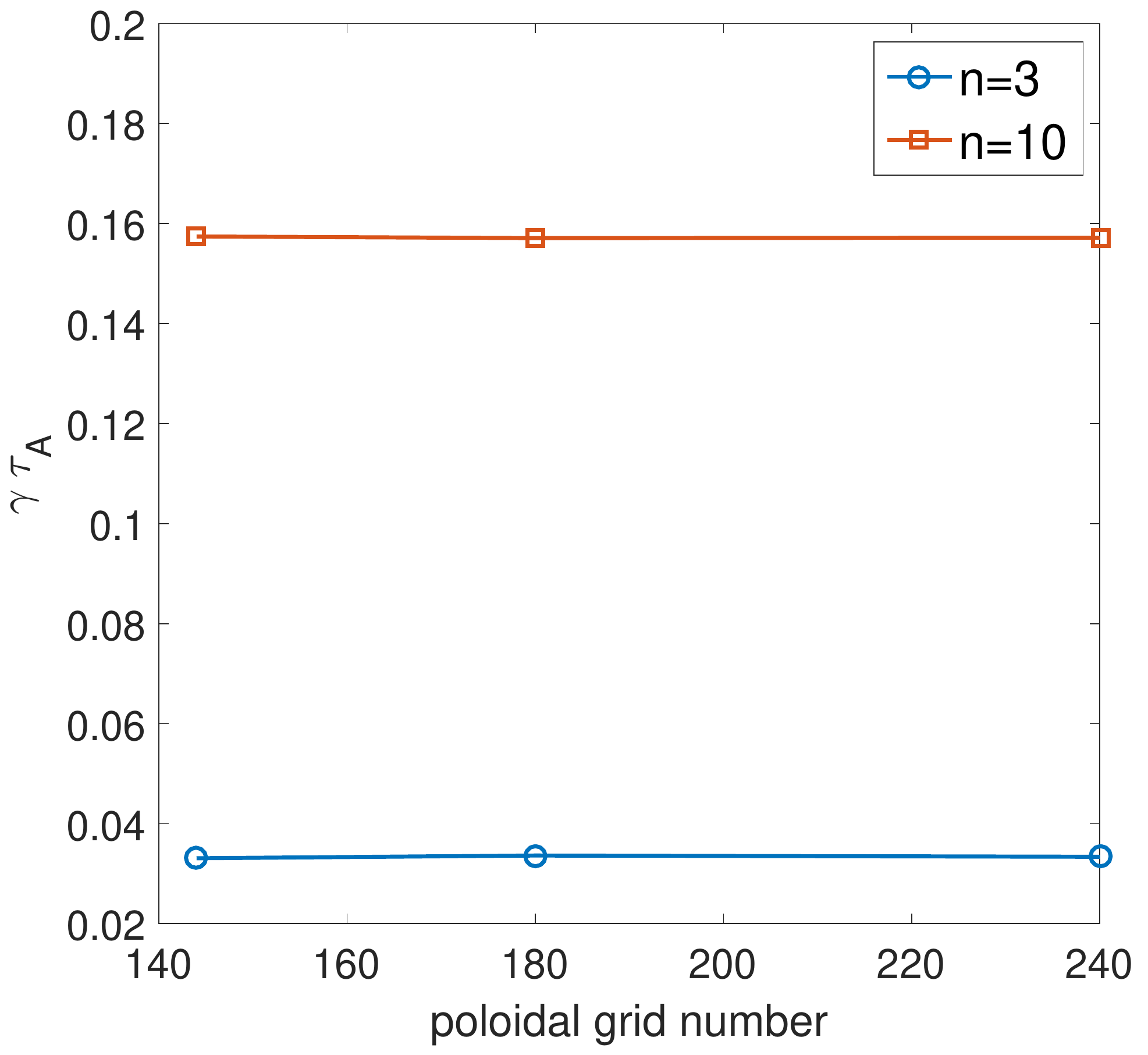}}
~~\subfloat[]{\includegraphics[width=7.5cm]{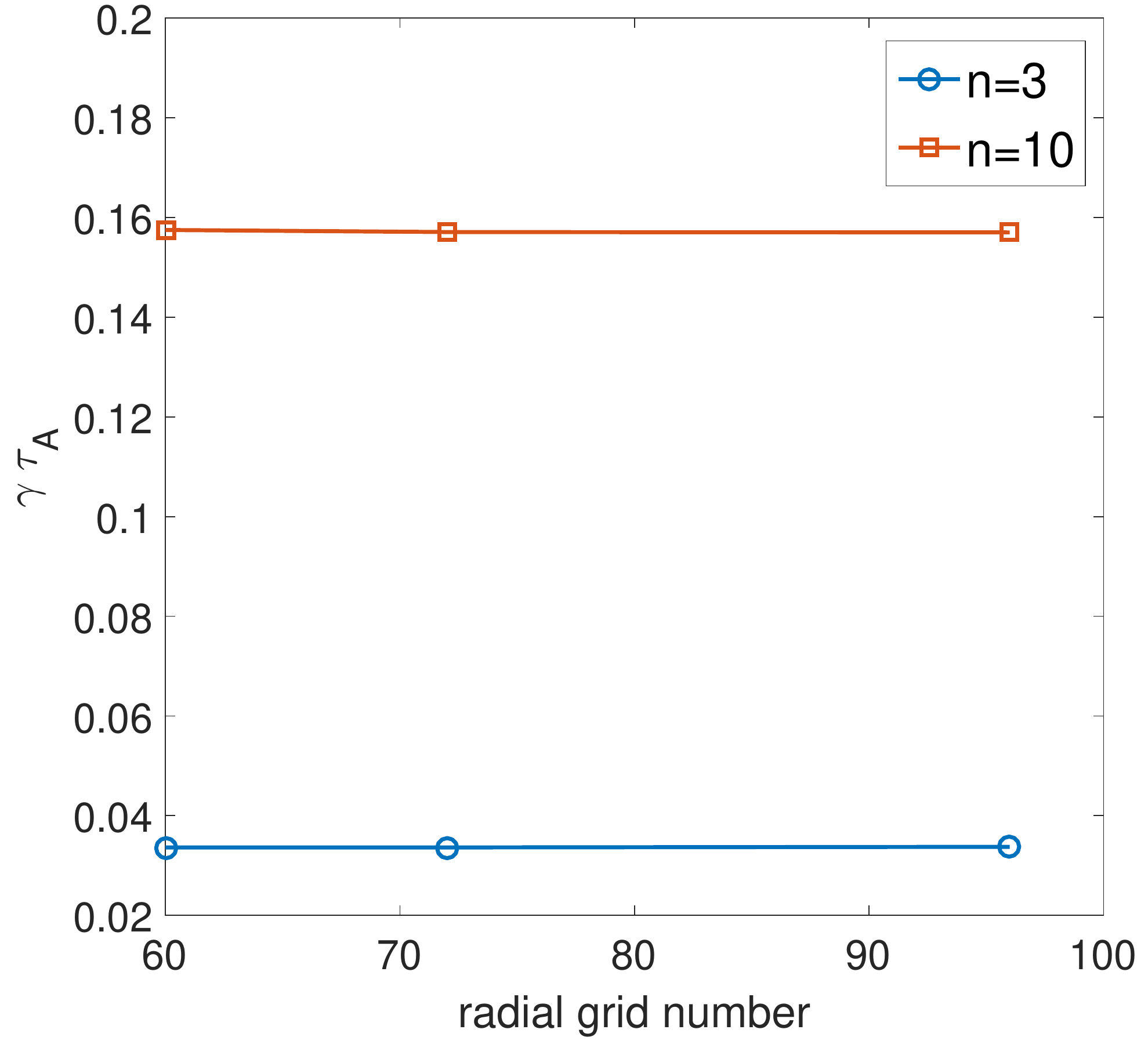}}\\
~~~~~~~~~~~~~~~~\\
~~~~~~~~~~~~~~~~\\
\subfloat[]{\includegraphics[width=7.5cm]{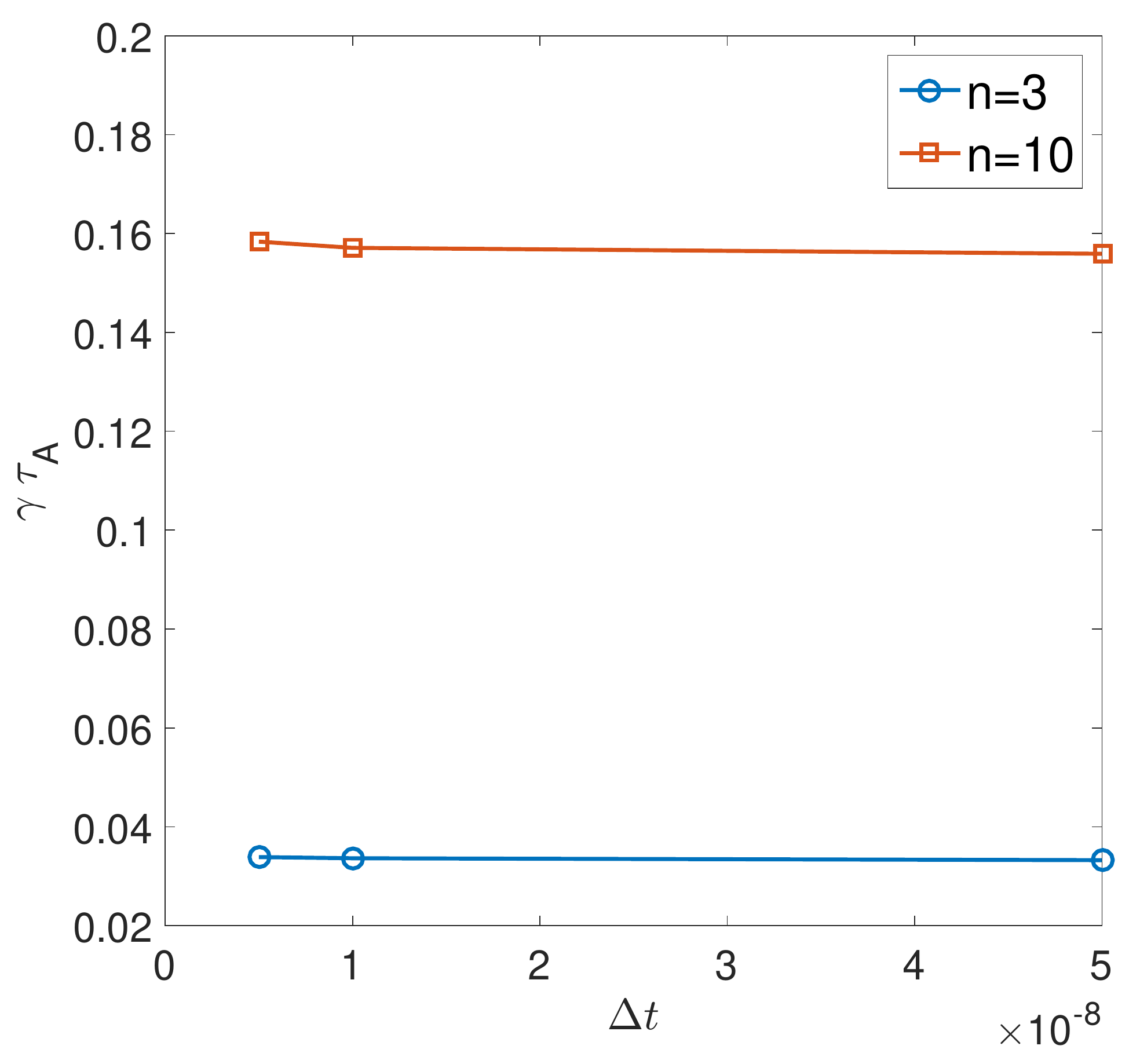}}
~~\subfloat[]{\includegraphics[width=7.5cm]{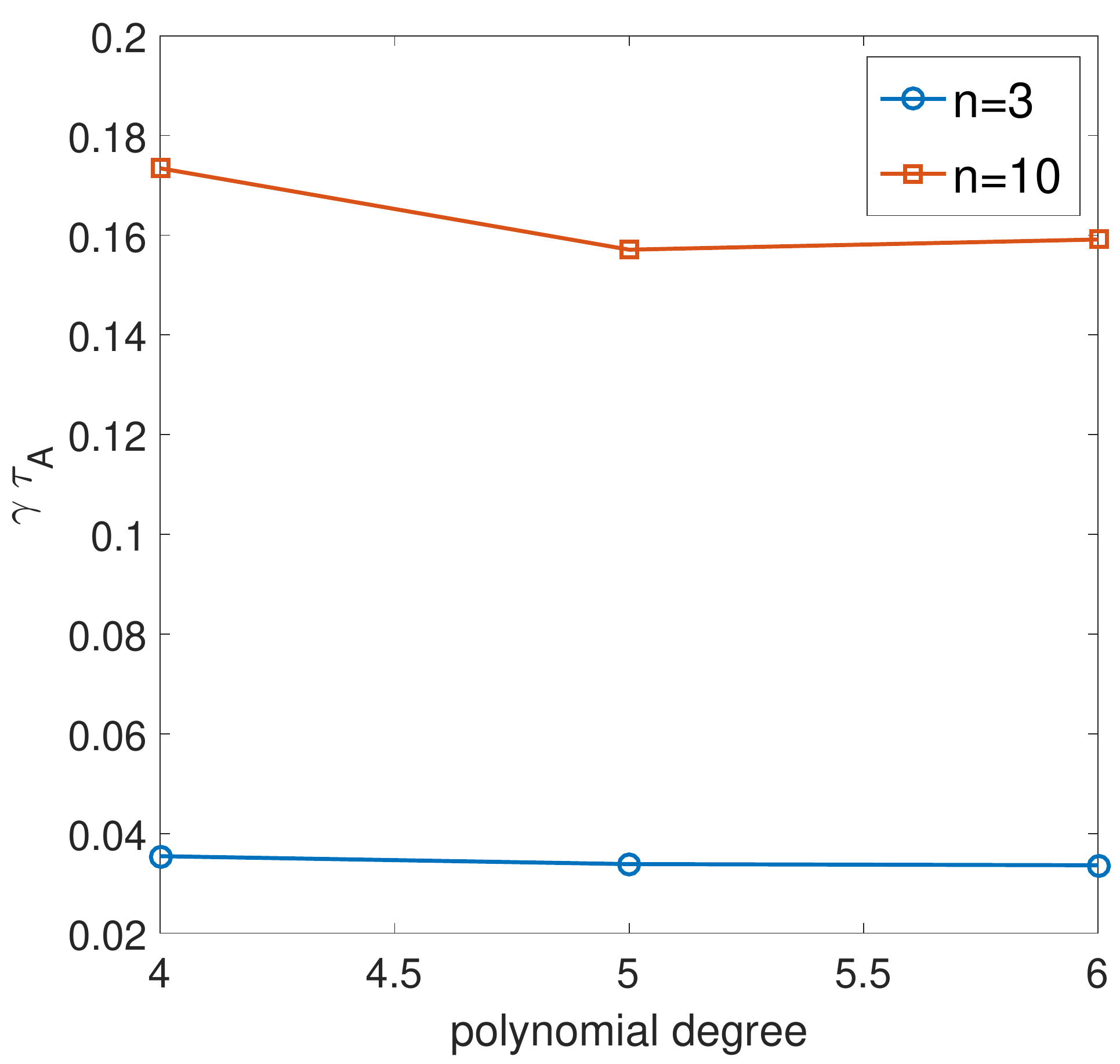}}\\
\caption{\label{con} Numerical convergence has been shown for poloidal grid number (a), radial grid number (b), time step (c) and polynomial degree (d). Two modes $n=3,10$
have been picked up for checking and found to have good convergence.}
\end{figure}
\end{document}